\documentclass[prd,aps,twocolumn,superscriptaddress,nofootinbib,preprintnumbers,floatfix]{revtex4-1}

\usepackage[bookmarks=false]{hyperref}
\usepackage{amsmath}
\usepackage{graphicx}
\usepackage[small]{subfigure}
\usepackage{placeins}
\usepackage{xcolor}
\usepackage{slashed}

\newcommand{\eq}[1]{Eq.~\eqref{eq:#1}}
\newcommand{\eqs}[2]{Eqs.~\eqref{eq:#1} and \eqref{eq:#2}}

\renewcommand{\sec}[1]{Sec.~\ref{sec:#1}}

\newcommand{\subsec}[1]{Sec.~\ref{subsec:#1}}

\newcommand{\fig}[1]{Fig.~\ref{fig:#1}}
\newcommand{\figs}[2]{Figs.~\ref{fig:#1} and \ref{fig:#2}}

\newcommand{\abs}[1]{\lvert#1\rvert}
\newcommand{\ord}[1]{\mathcal{O}(#1)}

\newcommand{\df}{\mathrm{d}}

\newcommand{\eps}{\epsilon}

\newcommand{\cO}{{\mathcal O}}

\newcommand{\bn}{\bar{n}}

\newcommand{\GeV}{\,\mathrm{GeV}}
\newcommand{\TeV}{\,\mathrm{TeV}}

\newcommand{\nn}{\nonumber}

\newcommand{\Ecm}{E_\mathrm{cm}}

\newcommand{\nons}{\mathrm{nons}}

\newcommand{\sub}{\mathrm{sub}}
\newcommand{\LO}{\mathrm{LO}}
\newcommand{\hadcm}{\mathrm{had\,cm}}

\newcommand{\Tau}{\mathcal{T}}

\newcommand{\cut}{{\mathrm{cut}}}

\allowdisplaybreaks[4]

\begin{document}

%%%%%%%%%%%%%%%%%%%%%%%%%%%%%%%%%%%%%%%%%%%%%%%%%%%%%%%%%%%%%%%%%%%%%%%%%%%%%%%%
% Title page
%%%%%%%%%%%%%%%%%%%%%%%%%%%%%%%%%%%%%%%%%%%%%%%%%%%%%%%%%%%%%%%%%%%%%%%%%%%%%%%%

\preprint{\vbox{\hbox{MIT--CTP 4855}\hbox{DESY 16-229}}}

\title{Subleading Power Corrections for N-Jettiness Subtractions}

\author{Ian Moult}
\affiliation{Center for Theoretical Physics, Massachusetts Institute of Technology, Cambridge, MA 02139, USA\vspace{0.5ex}}
\affiliation{Berkeley Center for Theoretical Physics, University of California, Berkeley, CA 94720, USA\vspace{0.5ex}}
\affiliation{Theoretical Physics Group, Lawrence Berkeley National Laboratory, Berkeley, CA 94720, USA\vspace{0.5ex}}

\author{Lorena Rothen}
\affiliation{Theory Group, Deutsches Elektronen-Synchrotron (DESY), D-22607 Hamburg, Germany\vspace{0.5ex}}

\author{Iain W.~Stewart}
\affiliation{Center for Theoretical Physics, Massachusetts Institute of Technology, Cambridge, MA 02139, USA\vspace{0.5ex}}

\author{Frank J.~Tackmann}
\affiliation{Theory Group, Deutsches Elektronen-Synchrotron (DESY), D-22607 Hamburg, Germany\vspace{0.5ex}}

\author{Hua Xing Zhu}
\affiliation{Center for Theoretical Physics, Massachusetts Institute of Technology, Cambridge, MA 02139, USA\vspace{0.5ex}}

\date{December 1, 2016}

%%%%%%%%%%%%%%%%%%%%%%%%%%%%%%%%%%%%%%%%%%%%%%%%%%%%%%%%%%%%%%%%%%%%%%%%%%%%%%%%
\begin{abstract}

The $N$-jettiness observable $\Tau_N$ provides a way of describing the leading singular behavior of the $N$-jet cross section in the $\tau =\Tau_N/Q \to 0$ limit, where $Q$ is a hard interaction scale. We consider subleading power corrections  in the $\tau \ll 1$ expansion, and employ soft-collinear effective theory to obtain analytic results for the dominant $\alpha_s \tau \ln\tau$ and $\alpha_s^2 \tau\ln^3\tau$ subleading terms for thrust in $e^+e^-$ collisions and $0$-jettiness for $q\bar q$-initiated Drell-Yan-like processes at hadron colliders. These results can be used to significantly improve the numerical accuracy and stability of the $N$-jettiness subtraction technique for performing fixed-order calculations at NLO and NNLO. They reduce the size of missing power corrections in the subtractions by an order of magnitude.
We also point out that the precise definition of $N$-jettiness has an important impact on the size of the power corrections and thus the numerical accuracy of the subtractions. The sometimes employed
definition of $N$-jettiness in the hadronic center-of-mass frame suffers from power corrections that grow exponentially with rapidity, causing the power expansion to deteriorate away from central rapidity.  This degradation does not occur for the original $N$-jettiness definition, which explicitly accounts for the boost of the Born process relative to the frame of the hadronic collision, and has a well-behaved power expansion throughout the entire phase space. Integrated over rapidity, using this $N$-jettiness definition in the subtractions yields another order of magnitude improvement compared to employing the hadronic-frame definition.

\end{abstract}
%%%%%%%%%%%%%%%%%%%%%%%%%%%%%%%%%%%%%%%%%%%%%%%%%%%%%%%%%%%%%%%%%%%%%%%%%%%%%%%%

\maketitle

%%%%%%%%%%%%%%%%%%%%%%%%%%%%%%%%%%%%%%%%%%%%%%%%%%%%%%%%%%%%%%%%%%%%%%%%%%%%%%%%
\section{Introduction}
\label{sec:intro}
%%%%%%%%%%%%%%%%%%%%%%%%%%%%%%%%%%%%%%%%%%%%%%%%%%%%%%%%%%%%%%%%%%%%%%%%%%%%%%%%

Precision QCD calculations play an essential role at the Large Hadron Collider (LHC), both for interpreting the measurement of Standard Model parameters, as well as in searches for new physics. In many cases, calculations at next-to-next-to-leading order (NNLO) in perturbative QCD are required for an accurate description of kinematic distributions and to be competitive with the ever increasing experimental precision.

Higher-order calculations in perturbative QCD involve infrared (IR) singularities arising from both real and virtual radiation, which cancel in the final result for any infrared and collinear safe quantity. Practical calculations require some method to isolate and cancel these IR singularities.
At NLO, this is very well understood and the standard method is to use FKS~\cite{Frixione:1995ms,Frixione:1997np} or CS~\cite{Catani:1996jh, Catani:1996vz, Catani:2002hc} subtractions, which construct local subtraction terms that approximate the real-emission amplitude in the IR limit point-by-point and whose integral can be carried out analytically and is added to the virtual contributions, such that the real and virtual contributions are separately rendered finite. Significant work has been focused on extending these subtraction techniques to NNLO, where the singularity structure is more complicated \cite{Weinzierl:2003fx, Frixione:2004is, Somogyi:2005xz, Somogyi:2006da, Somogyi:2006db, Somogyi:2008fc, Aglietti:2008fe, Bolzoni:2009ye, Bolzoni:2010bt, DelDuca:2013kw, Somogyi:2013yk, DelDuca:2015zqa, GehrmannDeRidder:2004tv, GehrmannDeRidder:2005cm, Daleo:2006xa, Daleo:2009yj, Glover:2010im, Boughezal:2010mc, Gehrmann:2011wi, GehrmannDeRidder:2012ja, Currie:2013vh, Currie:2013dwa, Anastasiou:2003gr, Binoth:2004jv, Czakon:2010td, Czakon:2011ve, Boughezal:2011jf, Brucherseifer:2013iv, Boughezal:2013uia, Czakon:2013goa, Czakon:2014oma, Boughezal:2015dra,DelDuca:2016csb,DelDuca:2016ily}, resulting in several approaches which have been succesfully applied to NNLO calculations with colored particles in the final state \cite{GehrmannDeRidder:2005cm,Czakon:2010td,Boughezal:2011jf,Czakon:2014oma}.

An alternative to fully local point-by-point subtractions is to use a physical jet-resolution variable to control the IR behaviour and construct suitable subtraction terms. This idea was originally applied to color-singlet production using the transverse momentum of the leptonic final state as a resolution variable \cite{Catani:2007vq}, and has also been used for top quark decays \cite{Gao:2012ja} using inclusive jet mass, and for $e^+e^- \to t\bar{t}$ using radiation energy \cite{Gao:2014nva}. Since all IR singular contributions are projected onto a single variable or dimension, such physical subtractions are intrinsically nonlocal, which may result in slower numerical convergence. Their key advantage is that by using a physical observable the subtraction terms are equivalent to the singular limits of a physical cross section, whose IR-singular structure is typically much easier to understand. Another benefit is that they allow one to directly reuse the existing NLO calculations for the corresponding Born$+1$-jet process. They are also conceptually straightforward to extend to even higher orders.

Recently a general subtraction framework based on the resolution variable $N$-jettiness $\Tau_N$~\cite{Stewart:2010tn} was proposed~\cite{Boughezal:2015dva, Boughezal:2015aha, Gaunt:2015pea}. It is applicable for an arbitrary number of jets in the final state. As explained in detail in Ref.~\cite{Gaunt:2015pea}, $N$-jettiness subtractions can be implemented either as differential subtractions in $\Tau_N$ or as global subtractions, which amounts to a phase-space slicing. The differential subtractions are effectively the basis of the \textsc{Geneva} method to match resummed NNLO calculation with parton showers~\cite{Alioli:2012fc,Alioli:2015toa}. Implemented as a global subtraction, they have been applied to calculate $W/Z/H+$ jet at NNLO~\cite{Boughezal:2015dva, Boughezal:2015aha, Boughezal:2015ded, Boughezal:2016dtm}, and have been implemented in MCFM for color-singlet production~\cite{Campbell:2016jau, Campbell:2016yrh, Boughezal:2016wmq}. They have also been applied to single-inclusive jet production in $ep$ collisions \cite{Abelof:2016pby}.

$N$-jettiness subtractions are based on parametrizing the phase space by the $N$-jettiness resolution variable, which is designed to vanish for an $N$-jet Born configuration. Explicitly, a generic $N$-jet cross section $\sigma(X)$, defined with Born level measurements and cuts $X$ can be written in terms of the integral of the corresponding differential cross section $\df\sigma(X)/\df\Tau_N$ as
%%%
\begin{align}
\sigma(X, \Tau_\cut) &\equiv \int^{\Tau_\cut}\!\df\Tau_N\, \frac{\df\sigma(X)}{\df\Tau_N}
\,, \nn \\
\sigma(X)
&= \sigma(X, \Tau_\cut) + \int_{\Tau_\cut} \df\Tau_N  \frac{\df\sigma(X)}{\df\Tau_N}
\,.\end{align}
%%%
From here on we will suppress the dependence on $X$.
We can now add and subtract a subtraction term $\df\sigma^\sub/\df\Tau_N$ to obtain
%%%
\begin{align} \label{eq:nsub_master}
\sigma
&= \bigl[\sigma(\Tau_\cut) - \sigma^\sub(\Tau_\cut) \bigr]
+ \sigma^\sub(\Tau_{\rm off})
\nn \\ & \quad
+ \int_{\Tau_\cut}\! \df\Tau_N \biggl[\frac{\df\sigma}{\df\Tau_N}
- \frac{\df\sigma^\sub}{\df\Tau_N} \theta(\Tau < \Tau_{\rm off}) \biggr]
\nn \\[0.5em]
&= \sigma^\sub(\Tau_\cut) + \int_{\Tau_\cut}\! \df\Tau_N \frac{\df\sigma}{\df\Tau_N}
+ \bigl[\sigma(\Tau_\cut) - \sigma^\sub(\Tau_\cut) \bigr]
\nn \\[0.5em]
&\equiv \sigma^\sub(\Tau_\cut) + \int_{\Tau_\cut}\! \df\Tau_N \frac{\df\sigma}{\df\Tau_N}
+ \Delta \sigma(\Tau_\cut)
\,,\end{align}
%%%
where in the last line we define $\Delta\sigma(\Tau_\cut)$. 
The value of $\Tau_{\rm off}$ is arbitrary and determines the range over which the subtraction acts differentially in $\Tau_N$.
In the last two lines we have set $\Tau^{\rm off} = \Tau_\cut$. This reduces the subtraction to a global phase-space slicing which is our focus here.

For $\Tau_N > \Tau_\cut$ an additional emission off of the $N$-jet Born configuration must be present and the calculation reduces to an $N+1$-jet perturbative calculation at one lower perturbative order. For NNLO calculations, the integral over $\Tau_N > \Tau_\cut$ can thus be obtained from an existing NLO calculation.
On the other hand, for $\Tau_N<\Tau_\cut$ the cross section is dominated by singular emissions from the Born configuration and can be approximated by $\sigma^\sub(X)$, which by construction reproduces its singular behaviour for $\Tau_N\to 0$. Hence, for sufficiently small $\Tau_\cut$ the difference $\Delta\sigma =\sigma(\Tau_\cut) - \sigma^\sub(\Tau_\cut)$ in \eq{nsub_master} scales as $\Tau_\cut$ and can be neglected.

In the limit of small $\Tau_N$, the cross section can be expanded in powers of $\tau_N=\Tau_N/Q$, where $Q$ is a typical hard scale inserted to make $\tau_N$ dimensionless, as
%%%
\begin{align}
\frac{\df \sigma}{\df\tau_N}
&= \frac{\df\sigma^{(0)}}{\df\tau_N} + \frac{\df\sigma^{(2)}}{\df\tau_N}+ \frac{\df\sigma^{(4)}}{\df\tau_N} + \dotsb
\,, \\ \nn
\sigma(\tau_\cut)
&= \sigma^{(0)}(\tau_\cut) + \sigma^{(2)}(\tau_\cut) + \sigma^{(4)}(\tau_\cut) + \dotsb
\,.\end{align}
%%%
Here, $\df\sigma^{(0)}/\df\tau_N$ and $\sigma^{(0)}(\tau_\cut)$ contain the leading-power (singular) terms which have the scaling
%%%
\begin{align}
\frac{\df\sigma^{(0)}}{\df\tau_N}
&\sim \delta(\tau_N)+ \biggl[\frac{ \ord{1} }{\tau_N}\biggr]_+
\,, \nn \\
\sigma^{(0)}(\tau_\cut) &\sim \ord{1}
\,.\end{align}
%%%
The ${\cal O}(1)$ factors include powers of $\ln\tau_N$ and $\ln\tau_\cut$ respectively. 
As indicated, the singular terms in the spectrum are divergent for $\tau_N\to 0$ and
are written in terms of distributions, which encode the cancellation
of real and virtual IR divergences. The subtraction terms in \eq{nsub_master}
must be equivalent to these leading-power terms, up to possible power-suppressed terms,
%%%
\begin{equation}
\sigma^\sub(\Tau_\cut) = \sigma^{(0)}(\tau_\cut = \Tau_\cut/Q)\, [1 + \ord{\tau_\cut}]
\,.\end{equation}
%%%

The $\df\sigma^{(2k)}/\df\tau_N$ and $\sigma^{(2k)}(\tau_\cut)$ terms with $k \geq 1$ are referred to as power corrections and contain the contributions that are suppressed by powers of $\tau_N$ relative to the leading-power terms,
%%%
\begin{align}\label{eq:scaling_lam2}
\tau_N \frac{\df\sigma^{(2k)}}{\df\tau_N} &\sim \ord{\tau_N^k}
\,, \quad
\sigma^{(2k)}(\tau_\cut) \sim \ord{\tau_\cut^k}
\,.\end{align}
%%%
The terms with $k = 1$ are the next-to-leading power (NLP) contributions. Note that the counting used here follows the standard power counting in soft-collinear effective theory (SCET), where the power counting parameter is $\lambda \sim \sqrt{\tau_N}$.
The power corrections have at most integrable divergences for $\tau_N\to 0$, i.e.\ they do not depend on purely virtual corrections to the Born process and can thus be computed from the corresponding $N+1$-jet process. (This is also why they do not necessarily have to be included in the subtractions.)

By using a physical observable like $\Tau_N$, the leading singular terms $\df\sigma^{(0)}/\df\tau_N$ can be computed using a factorization theorem for the cross section~\cite{Stewart:2009yx, Stewart:2010tn} derived in SCET~\cite{Bauer:2000ew, Bauer:2000yr, Bauer:2001ct, Bauer:2001yt, Bauer:2002nz}, in terms of universal jet, soft, and beam functions describing the soft and collinear limits of QCD. This provides explicit analytic control over the infrared divergent contributions. The required ingredients to obtain $\sigma^{(0)}(\tau_\cut)$ to NNLO are 
the NNLO jet \cite{Becher:2006qw, Becher:2010pd} and beam \cite{Gaunt:2014xga, Gaunt:2014cfa} functions, which are fully known, as well as the soft function which is fully known to NNLO for two external partons \cite{Kelley:2011ng, Monni:2011gb, Hornig:2011iu, Kang:2015moa}. The soft function for arbitrary $N$ is currently known to NLO~\cite{Jouttenus:2011wh}. The soft function for three external partons, as relevant for color singlet plus jet production at the LHC, was computed numerically at NNLO in Ref.~\cite{Boughezal:2015eha} and for a massive third parton in Ref.~\cite{Li:2016tvb}.

\begin{figure*}[t]
\includegraphics[width=\columnwidth]{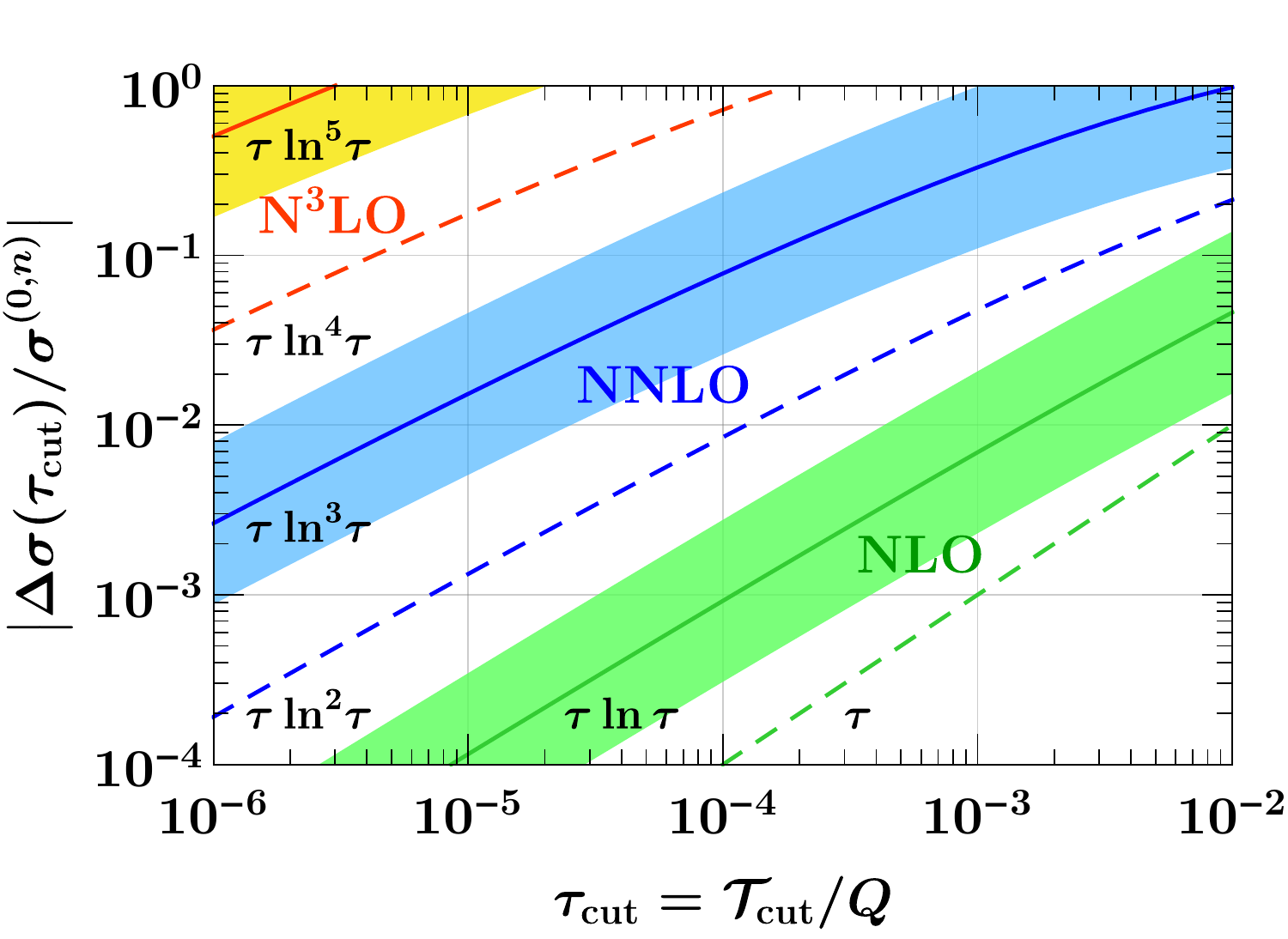}%
\hfill%
\includegraphics[width=\columnwidth]{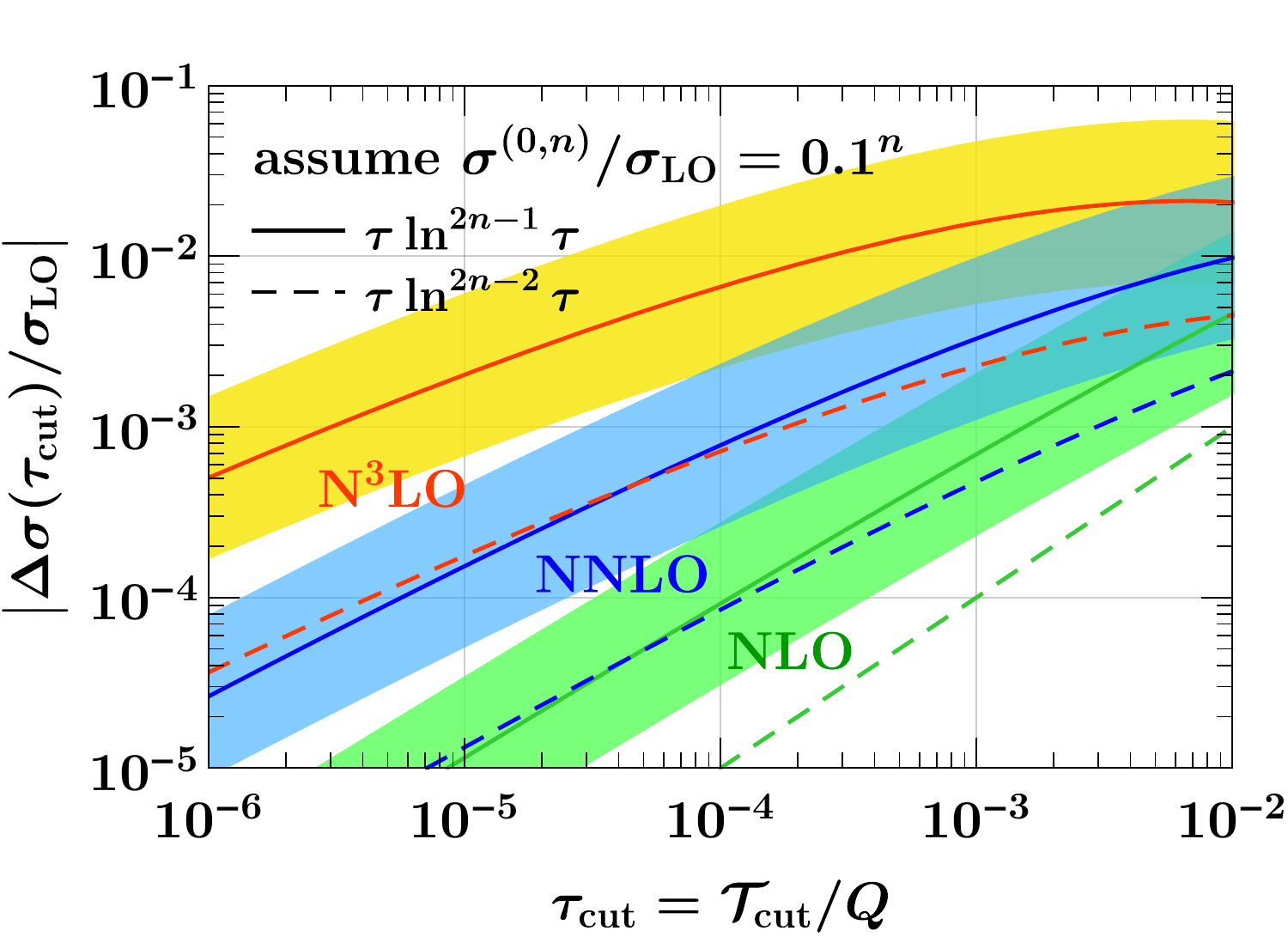}%
\caption{Estimate of the missing power corrections $\Delta\sigma(\tau_\cut)$ below $\tau_\cut$ for the NLO (green), NNLO (blue), and N$^3$LO (orange) contributions without including (solid) and including (dashed) the leading power correction in the subtractions. On the left, the estimate is relative to the full N$^n$LO contribution itself, on the right relative to the LO cross section. The bands show a factor of three variation in the estimate around the solid lines. A similar variation should be considered to apply to the dashed lines, but for simplicity is not shown.}
\label{fig:scaling}
\end{figure*}

When employing the subtraction method,
the size of the neglected contributions below $\Tau_\cut$ determine the
error one makes in the calculated cross section, which is given by
%%%
\begin{equation}
\Delta\sigma(\tau_\cut)
= \sigma(\tau_\cut) - \sigma^\sub(\tau_\cut)
= \sigma^{(2)}(\tau_\cut) + \dotsb
\,,\end{equation}
%%%
and is thus controlled by the NLP corrections $\sigma^{(2)}(\tau_\cut) \sim \ord{\tau_\cut}$.
As discussed in detail in Ref.~\cite{Gaunt:2015pea}, the expected error can thus be estimated
based on the perturbative structure of the neglected NLP corrections. Writing the perturbative expansion in $\alpha_s$ as
%%%
\begin{align}
\frac{\df\sigma^{(k)}}{\df\tau_N}
&= \sum_{n=0} \frac{\df\sigma^{(k,n)}}{\df\tau_N} \Bigl( \frac{\alpha_s}{4\pi} \Bigr)^n
\,,\end{align}
%%%
the perturbative structure of the dominant power corrections is given by
%%%
\begin{align}
\tau_N  \frac{\df\sigma^{(2,n)}}{\df\tau_N}
&= \tau_N \sum_{m=0}^{2n-1} \ C^{(2,n)}_m \ln^m \tau_N
\,,\nn \\
\sigma^{(2,n)}(\tau_\cut)
&=  \tau_\cut \sum_{m=0}^{2n-1}\ A^{(2,n)}_m \ln^{m} \tau_\cut
\,,\end{align}
%%%
where the $A^{(2,n)}_m$ coefficients are straightforwardly related to the $C^{(2,n)}_{m'}$ coefficients by integration.

Hence, the dominant behaviour of the power correction is $\alpha_s \,\tau_\cut \ln\tau_\cut$ at NLO and
$\alpha^2_s\, \tau_\cut \ln^3 \tau_\cut$ at NNLO, and so forth.
While these corrections vanish in the limit $\tau_\cut \to 0$, they do so slower and slower at higher orders due to the strong logarithmic enhancement, requiring very small values of $\tau_\cut$ to be used in the subtractions.
In \fig{scaling} we show an estimate of the error due to missing power corrections $\Delta\sigma(\tau_\cut)$ as a function of $\tau_\cut$, based on the form of their leading-logarithmic term relative to the leading-power $\alpha_s^n$ coefficient $\sigma^{(0,n)}$ (on the left) and relative to the LO cross section (on the right, assuming a $10\%$ correction at each order in $\alpha_s$).
The bands show a variation of the estimate by a factor of three.
We see that for a fixed value of the cutoff, the size of the missing terms grows rapidly with the loop order.
On the other hand, in practice reducing $\tau_\cut$ comes at the price of a reduced numerical stability in the NLO $N+1$-jet calculation and quickly increasing computational time required to obtain small statistical uncertainties in the Monte-Carlo integration. The typical values used in current implementations are in the $\tau_\cut \simeq 10^{-3}$ to $10^{-4}$ range.%

A possibility to greatly improve the numerical stability of the subtraction, which was already put forth in Ref.~\cite{Gaunt:2015pea}, is to explicitly compute the dominant power corrections and include them in the subtractions. The dashed lines in \fig{scaling} show an estimate of the error $\Delta(\sigma_\cut)$ when including the leading-logarithmic power correction, $C^{(2,k)}_{2k-1}$, in the subtractions $\sigma^\sub(\tau_\cut)$. Based on this simple estimate, for small values of $\tau_\cut$, this can reduce the error by about an order of magnitude, or equivalently for fixed error allow one to raise $\tau_\cut$ by an order of magnitude.
This trend continues with each logarithm that is added to the subtraction. Therefore, both the numerical stability and accuracy of the subtraction can be significantly improved by computing the power corrections. Note that the power corrections would give an $\mathcal{O}(1)$ error to the N$^3$LO coefficient even for very small values of $\tau_\cut$, and therefore to make the application feasible at this order it will be absolutely essential to include the leading-power corrections in the subtractions.

The goal of this paper is to analytically calculate the leading-logarithmic (LL) terms $C^{(2,1)}_{1}$ and $C^{(2,2)}_{3}$ at NLO and NNLO for $0$-jettiness, which is equivalent to beam thrust~\cite{Stewart:2009yx, Stewart:2010pd}, for $q \bar q$-initiated Drell-Yan-like processes. Like at leading power, this is made feasible by virtue of the fact that $\Tau_N$ is a physical observable. Our calculation will be performed in SCET, which features a systematic power expansion. To ensure that we have identified all sources for the power corrections we exploit the recently determined complete basis of hard scattering operators for $e^+e^-\to $ dijets and Drell-Yan from $q\bar q$ annihilation~\cite{Feige:2017zci}.  We will emphasize more generally how SCET can be used to analytically compute power corrections for physical resolution variables.
By calculating the LL terms exactly, we will also be able to numerically extract the next-to-leading logarithmic (NLL) contributions from the full fixed-order results. We study in detail their effect on improving $N$-jettiness subtractions.
Our numerical results also confirm the naive scaling estimates shown in \fig{scaling}.

We will also highlight an important point regarding the precise definition used for $\Tau_N$, which can strongly impact the size of the power corrections as a function of the Born phase space. We show that the definition of $\Tau_N$ in the hadronic (lab) frame utilized in some applications of $N$-jettiness subtractions, suffers form exponentially enhanced power corrections at large rapidities, leading to a deterioration of the power expansion and thus the utility of the $N$-jettiness subtractions there. This is avoided by employing the original and more natural definition of $N$-jettiness that incorporates the boost of the Born system relative to the frame of the hadronic collisions. Compared to the hadronic definition, this alone leads to a reduction of the power corrections by an order of magnitude, as well as stable behavior throughout the entire phase space.

In this work we focus on the main ideas, results, and analysis, leaving a detailed exposition of the SCET based organization and calculational techniques to a future publication.
The remainder of this paper is organized as follows. In \sec{calc}, we present our calculation. We derive general consistency relations for the cancellation of infrared poles at subleading power, which can both be used as a check on the calculation, as well as a simplification. We then calculate the terms
\begin{align}
C^{(2,1)}_{1} \, \frac{\alpha_s}{4\pi}\, \tau\ln\tau
\,,\qquad
C^{(2,2)}_{3} \, \Bigl(\frac{\alpha_s}{4\pi} \Bigr)^2\, \tau\ln^3 \tau \,,
\end{align}   
for both thrust in $e^+e^-$ and $0$-jettiness in $pp$ collisions. We present numerical results at NLO and NNLO and perform a detailed comparison with the numerical fixed-order results for $Z+1$ jet from MCFM~\cite{Campbell:1999ah, Campbell:2010ff, Campbell:2015qma, Boughezal:2016wmq}. In \sec{discuss}, we discuss the dependence of the power corrections on the $N$-jettiness definition, and discuss the implications of our calculations for future use of the $N$-jettiness subtractions. We conclude in \sec{conclusions}.

%%%%%%%%%%%%%%%%%%%%%%%%%%%%%%%%%%%%%%%%%%%%%%%%%%%%%%%%%%%%%%%%%%%%%%%%%%%%%%%%
\section{Calculation}
\label{sec:calc}
%%%%%%%%%%%%%%%%%%%%%%%%%%%%%%%%%%%%%%%%%%%%%%%%%%%%%%%%%%%%%%%%%%%%%%%%%%%%%%%%

In this section, we present our calculation of the $C^{(2,1)}_{1}$ and $C^{(2,2)}_{3}$ coefficients  for $0$-jettiness. We begin in \subsec{constraint} by deriving general consistency constraints on our calculation arising from the cancellation of $1/\epsilon$ poles. In \subsec{thrust} we calculate the coefficients $C^{(2,1)}_{1}$ and $C^{(2,2)}_{3}$ for thrust in $e^+e^-\to$ dijets, which removes any complications related to the parton distribution functions (PDFs). Finally, in \subsec{beam_thrust} we cross our results from thrust to the case of $0$-jettiness, and present numerical results.

We organize our calculation using SCET~\cite{Bauer:2000ew, Bauer:2000yr, Bauer:2001ct, Bauer:2001yt, Bauer:2002nz}, which is an effective field theory of QCD describing the interactions of collinear and soft particles in the presence of a hard interaction. It is formulated as an expansion about the soft and collinear limits in a power counting parameter $\lambda$, which in our case corresponds to $\lambda^2 \sim \tau_N$. SCET is explicitly constructed to maintain manifest power counting in all stages of a calculation. In particular, all fields and Lagrangians are assigned a definite power counting~\cite{Bauer:2001ct}. The effective theory provides a natural organization of the different sources of power corrections. It also allows for the use of symmetries, such as reparametrization invariance (RPI) \cite{Manohar:2002fd,Chay:2002vy}, to relate certain power-suppressed contributions to leading-power results. It has been used to study factorization theorems at subleading power for $B$ decays \cite{Beneke:2002ni, Beneke:2004in, Lee:2004ja, Hill:2004if, Bosch:2004cb, Beneke:2004rc, Paz:2009ut, Benzke:2010js}, to study subleading soft limits at the amplitude level \cite{Larkoski:2014bxa}, and subleading factorization and resummation of the event shape thrust in $e^+e^-$ \cite{Freedman:2013vya, Freedman:2014uta}.

Power corrections arise from three sources in the effective theory:  universal subleading Lagrangian insertions, which correct the dynamics of the soft and collinear fields, subleading-power hard scattering operators, which describe local corrections to the hard scattering vertex, and subleading terms in the expansion of the measurement functions and phase space. Power-suppressed Lagrangians have been studied in the literature~\cite{Beneke:2002ni,Chay:2002vy,Manohar:2002fd,Pirjol:2002km,Beneke:2002ph,Bauer:2003mga}, and the SCET Lagrangian is known to ${\cal O}(\lambda^2)$ \cite{Bauer:2003mga}. For the processes we consider here, $e^+e^-\to $ dijets and Drell-Yan from $q\bar q$ annhilation, a complete basis of SCET hard scattering operators was presented in Ref.~\cite{Feige:2017zci}. The subleading power expansion for the measurement function for thrust has also been derived, originally in Ref.~\cite{Freedman:2013vya} with a different formalism than the one employed here, and also in Ref.~\cite{Feige:2017zci}.

Our calculation is carried out by using SCET as a means to organize the fixed-order calculation for the perturbative power corrections, and identify the most singular terms. For the observables discussed here this involves considering real-emission diagrams involving soft and/or collinear particles, as well as virtual corrections from soft and collinear loops, plus short-distance hard loops. The hard, collinear, or soft loops or emissions each only involve a single scale.
For this purpose, we do not need to make use of factorization theorems for these power corrections.

%===============================================================================
\subsection{General constraints from consistency}
\label{subsec:constraint}
%===============================================================================

We first discuss general constraints arising from the cancellation of $1/\epsilon$ poles on the subleading power cross section. In SCET these poles are ultraviolet in origin and arise from scale separation in the hard, collinear, and soft regions. (From a full theory point of view these poles are tracking infrared scales, and hence cancel because the subleading power cross section is free of non-trivial infrared divergences.)  Here we consider a generic dimensionless $\mathrm{SCET}_{\mathrm I}$ observable $\tau$, which we will later take to be thrust or $0$-jettiness in our explicit calculations. The consistency relations are, however, generic. If we compute the cross section using bare contributions from both the hard Wilson coefficients and collinear and soft phase space and loop integrals, then the $\cO(\tau)$ correction to the cross section has the following general expression,
%%%
\begin{align}\label{eq:constraint_setup}
\frac{\df\sigma^{(2,n)}}{\df\tau}
  = & \sum_{\kappa}\sum_{i=0}^{2n-1} \frac{c_{\kappa,i}}{\epsilon^i} \left( \frac{\mu^{2n}}{Q^{2n} \tau^{m(\kappa)}}   \right)^\epsilon
\nonumber
\\
& + \sum_{\gamma}\sum^{2n-2}_{i=0} \frac{d_{\gamma,i}}{\epsilon^i} 
\left(
\frac{\mu^{2(n-1)}}{Q^{2(n-1)} \tau^{m(\gamma)}}   \right)^\epsilon
\nonumber
\\
& + \dots
\,.\end{align}
%%%
Here $\kappa$ and $\gamma$ label the scalings obtained from the contributing particles, i.e., hard, collinear, or soft, and $m(\kappa)\geq 1$ is an integer. For example, at one loop ($n=1$) there is a single additional particle, which is either
%%%
\begin{align} \label{eq:classes1}
\text{soft:} \qquad &\kappa=s\,, \qquad m(\kappa) =2\,,\nn \\
\text{collinear:} \qquad &\kappa=c\,, \qquad m(\kappa)=1 \,,
\end{align}
%%%
while at two loops ($n=2$), the possible contributions are
%%%
\begin{align} \label{eq:classes2}
\text{hard-collinear:}\qquad  &\kappa = hc\,, \qquad m(\kappa) =1\,,\nn \\
 \text{hard-soft:} \qquad &\kappa= hs\,, \qquad m(\kappa) =2\,,\nn \\
\text{ collinear-collinear:} \qquad & \kappa=cc\,, \qquad m(\kappa) =2\,,\nn \\
 \text{collinear-soft:} \qquad & \kappa= cs\,, \qquad m(\kappa) =3\,,\nn \\
 \text{soft-soft:}  \qquad & \kappa= ss\,,\qquad m(\kappa) =4\,.
 \end{align}
 %%%
In general the number of terms in $\kappa$ indicates the loop order. 
 The extension to determine scalings at higher loops should be obvious.
 In the first line of \eq{constraint_setup} the $c_{\kappa,i}$ are the coefficients of the poles for each different contribution. Starting from the second line the contribution from UV renormalization, and collinear PDF renormalization in the case of hadronic collisions, is taken into account. Here $\gamma$ is similar to $\kappa$ but with one less $h$, $c$, or $s$ in it at each order in $\alpha_s$. For example, for $0$-jettiness at a hadron collider, schematically we have
\begin{align}
  \label{eq:counter_term}
  d_{\gamma,2} = - \beta_0 c_{\gamma,1} +  \sum_f (P^{(0)}\otimes c_{\gamma,1})_f\,,\quad \gamma=s,c\,,
\end{align}
where $\beta_0$ is the one-loop beta function coefficient, and $P^{(0)}$ is the LO splitting function with appropriate partonic flavor, and the $\sum_f$ denotes summation over different flavor
combinations and incoming legs. We should point out that the $d_{\gamma,i}$ 
coefficients, as well as higher-order terms shown by the ellipses in \eq{constraint_setup} are completely fixed by the known beta function, splitting functions, and lower-order terms.

At subleading power there are no purely hard corrections, which means $m(\kappa)\geq 1$.
Demanding that the pole terms cancel in \eq{constraint_setup} gives rise to a number of constraints on the coefficients $c_{\kappa,i}$, which can be exploited to drastically simplify the calculation, as well as to provide cross checks on the result.

At one loop, where there is either a single soft or collinear emission in SCET, the cancellation of the pole terms in \eq{constraint_setup} yields the simple constraint
%%%
\begin{equation}\label{eq:one_loop_constraint}
c_{s,1}=-c_{c,1}
\,.\end{equation}
%%%
At two loops we find the following nontrivial constraints on the coefficients
%%%
\begin{align}\label{eq:constraints_summary}
c_{hc,3} &= \frac{c_{cs,3}}{3} = -c_{ss,3}= - \frac{1}{3} (c_{hs,3}+c_{cc,3})
\,, \nn \\
c_{cs,2}  &= c_{hc,2}-2c_{ss,2}+d_{c,2}
\,,\\
c_{hs, 2} + c_{cc,2} &= -2 c_{hc,2} + c_{ss,2} - d_{c,2}
\,, \nn \\
c_{hs,1} + c_{cc,1} &= - (c_{cs,1} + c_{hc,1} + c_{ss,1} + d_{c,1} + d_{s,1})
\,,\nn\end{align}
%%%
which apply separately in each color channel. Note that we have applied the relation $d_{s,2} = - d_{c,2}$ which is a consequence of \eq{counter_term} and \eq{one_loop_constraint}.

These consistency relations allow us to reduce the number of unknown coefficients at two loops. 
We can express the two-loop result for the subleading power correction as
%%%
\begin{align}\label{eq:constraints_final}
\frac{\df\sigma^{(2,2)}}{\df\tau}
&= c_{hc,3} \ln^3 \tau
+ (c_{hc,2} + c_{ss,2} + d_{c,2}) \ln^2 \tau
\nn \\ & \quad
+ \left(-c_{cs, 1} + c_{hc, 1} - 2 c_{ss, 1} + d_{c, 1}    \right) \ln\tau
\nn\\ & \quad
+  d_{c, 2} \ln\frac{Q^2}{\mu^2} \ln\tau
+\text{const}
\,.\end{align}
%%%
Here we have chosen to write the result in terms of hard-collinear terms whenever possible, as these terms have the simplest phase-space integrals. For other applications,
other organizations may prove more convenient. Interestingly, for the $\ln^3 \tau$ term, on which we focus here, the consistency relations imply that we only need to calculate a single two-loop coefficient, $c_{hc,3}$ (alternatively $c_{ss,3}$ or $c_{cs,3}$). By calculating more coefficients the constraints of \eq{constraints_summary} can then provide a powerful check on our calculation.

Although the focus of this paper is on the leading-logarithmic term, $\ln^3 \tau$, the consistency relations also significantly reduce the number of unknown coefficients for the $\ln^2 \tau$ and $\ln \tau$ terms. Scale dependence first appears in the coefficient of $\ln\tau$, as expected.
In the future, we hope to exploit these relations to analytically compute the lower-order terms. We also believe that these relations may prove useful in the extension of our current calculation to the case of processes involving jets in the final state. In particular, the coefficient of the $\ln^3 \tau$ term is determined entirely by the hard-collinear contribution. Understanding the universality of the subleading collinear limits may therefore allow for calculations at subleading power to be extended straightforwardly to final states involving additional jets. 

%===============================================================================
\subsection{\boldmath $2$-jettiness in $e^+e^- \to$ jets (thrust)}
\label{subsec:thrust}
%===============================================================================

We begin by computing $2$-jettiness in $e^+e^-$ to jets. For massless partons this is equivalent to thrust \cite{Farhi:1977sg}, for which the exact one-loop result is known, and will provide a cross check on our results.
The thrust measurement function is defined by
%%%%
\begin{align}\label{eq:thrust_defn}
\tau=1- \text{max}_{\hat t} \frac{\sum_i |\hat t \cdot \vec p_i|}{\sum_i |\vec p_i|}\,.
\end{align}
%%%%

We focus on the leading-logarithmic terms in $\sigma^{(2,n)}$ at one and two loops, which scale as $\alpha_s \ln\tau$ and $\alpha^2_s \ln^3 \tau$, respectively. We will discuss in some detail the structure of the calculation at NLO, focusing on the different types of power corrections, and the cancellation of $1/\epsilon$ poles. We then use the result of \eq{constraints_final} to extend this calculation to NNLO.

%~~~~~~~~~~~~~~~~~~~~~~~~~~~~~~~~~~~~~~~~~~~~~~~~~~~~~~~~~~~~~~~~~~~~~~~~~~~~~~~
\subsubsection{Power Corrections at NLO}
%~~~~~~~~~~~~~~~~~~~~~~~~~~~~~~~~~~~~~~~~~~~~~~~~~~~~~~~~~~~~~~~~~~~~~~~~~~~~~~~

By studying contributions from the complete basis of SCET  operators and possible Lagrangian insertions, we  determine that there are four contributions to the leading logarithm at NLO that must be computed. 
These can be grouped into two categories, that separately exhibit the cancellation of $1/\eps$ poles
\begin{itemize}
\item Category 1: A gluon becomes collinear with a quark, or becomes soft.
\item Category 2: Two quarks becomes collinear, or a quark becomes soft.
\end{itemize}
Category 1 is of course familiar from the leading-power case, while Category 2 first appears at subleading power and will give rise to a $C_A$ color factor in the leading logarithm.  In the effective field theory organization, there are two classes of diagrams which contribute to each of these categories at NLO,  corresponding to a  soft or collinear emission as enumerated in \eq{classes1}. We discuss the contributions from the two categories in turn. We carry out our calculations in Feynman gauge, and note that the contributions from the categories and classes are individually gauge invariant.

\begin{figure}[t]
\subfigure[]{\includegraphics[width=0.49\columnwidth]{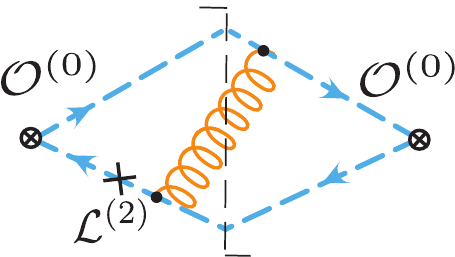}}
\hfill
\subfigure[]{\includegraphics[width=0.47\columnwidth]{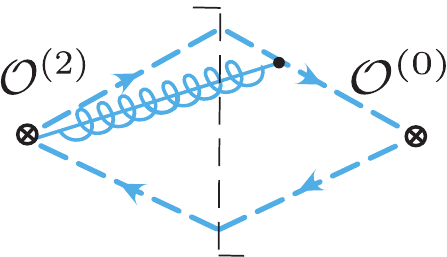}}
\caption{Representative NLO diagrams where a gluon becomes either soft (a) or collinear (b) with a quark. Collinear particles are shown in light blue, soft particles in orange. The cross represents a Lagrangian correction to the propagator, and the power suppression of the hard scattering operators and Lagrangian insertions is explicitly indicated.}
\label{fig:category1}
\end{figure}

First consider Category 1. The nonzero subleading power corrections in the case that a gluon becomes soft are reproduced by an SCET Lagrangian insertion correcting the collinear quark propagator (which is equivalent to the Low-Burnett-Kroll theorem \cite{Low:1958sn,Burnett:1967km} interfered with an eikonal emission, see Ref.~\cite{Larkoski:2014bxa}), while the subleading power corrections to the collinear
limit are reproduced by a power suppressed hard scattering operator. These both correspond to corrections to the amplitudes, and are illustrated in \fig{category1}, where the power suppression of the hard scattering operators, and Lagrangian insertions are indicated. The power suppression of a diagram is given by the sum of the power suppression of the Lagrangian insertions, and the hard scattering operators. Collinear particles are shown in light blue, while soft particles are shown in orange. Since both these limits exist at leading 
power, there are also potentially subleading power corrections from the expansion of the thrust measurement function, and the phase space. The 
subleading power expansion of the thrust measurement function was given in Refs.~\cite{Freedman:2013vya,Feige:2017zci}, and does not contribute a leading logarithmic divergence.
On the other hand, corrections to the phase space do give rise to a leading logarithmic divergence. In the effective field theory, we are free to choose the routing of 
residual momenta. Considering as an example the graph with a single soft gluon in \fig{category1}a, and let the momentum of the collinear particles be in the $n$ and $\bar n$ directions, where $n$ denotes a lightlike vector, and $\bar n$ its spatial conjugate. We can then route the $n$ component of the soft gluon's momentum through the $n$ collinear quark, the $\bar n$ component through the $\bar n$ collinear quark, and the perp component through the incoming off-shell propagator. This then trivially corrects the phase space integrals for the two collinear particles.

\begin{figure}[t]
\subfigure[]{\includegraphics[width=0.49\columnwidth]{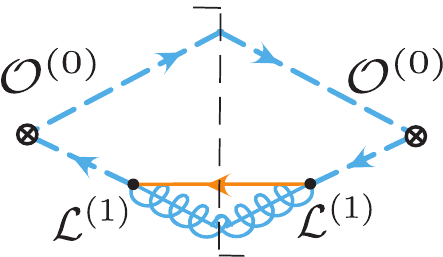}}
\hfill
\subfigure[]{\includegraphics[width=0.49\columnwidth]{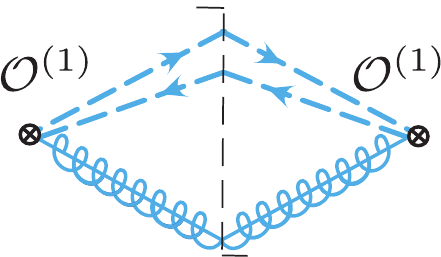}}
%%%
\caption{Representative NLO diagrams when a quark becomes either soft (a) or two quarks becomes collinear (b). The power suppression of the hard scattering operators and Lagrangian insertions is explicitly indicated.}
\label{fig:category2}
\end{figure}

At NLO, we find for Category 1
%%%
\begin{align}\label{eq:cat1_NLO}
\frac{1}{\sigma_0}\frac{\df\sigma_\mathrm{Cat.1}^{(2,1)}}{\df\tau}
&= 8C_F \biggl[ \biggl( \frac{1}{\epsilon}+\ln\frac{\mu^2}{Q^2 \tau} \biggr)
- \biggl( \frac{1}{\epsilon} +\ln\frac{\mu^2}{Q^2 \tau^2} \biggr) \biggr]
\nn \\
&= 8C_F \ln \tau
\,,\end{align}
%%%
where in the first equality we have separated the contributions from soft and collinear graphs.
This result can also be obtained by expanding the appropriate QCD amplitudes in the limit that the gluon becomes collinear or soft. Note that the $1/\eps$ poles cancel amongst the diagrams in this category, in agreement with the constraint of \eq{one_loop_constraint}. 
We also see the appearance of the characteristic soft scale $Q^2 \tau^2$ and collinear scale $Q^2\tau$ in the result in \eq{cat1_NLO}.

The analysis of the two classes of diagrams in Category 2 is a bit simpler, since in this case there are no corrections to the phase space. In the effective theory, the soft quark limit is reproduced by a subleading power Lagrangian insertion, while the limit of two collinear quarks is reproduced by a hard scattering operator. Representative diagrams are shown in \fig{category2}, and indicate the source of power suppression. 

The result for the Category 2 contributions is given by
%%%
\begin{align}
\frac{1}{\sigma_0}\frac{\df\sigma_\mathrm{Cat.2}^{(2,1)}}{\df\tau}
&= 4C_F \biggl[ -\biggl(\frac{1}{\epsilon} + \ln\frac{\mu^2}{Q^2 \tau} \biggr)
+ \biggl(\frac{1}{\epsilon} + \ln\frac{\mu^2}{Q^2 \tau^2}  \biggr) \biggr]
\nn \\
&= -4C_F \ln\tau
\,.\end{align}
%%%
We again see the cancellation of $1/\eps$ poles.
Also in this case, the result can be obtained by expanding the relevant QCD diagram in the limit of a quark going soft, or two quarks becoming collinear.

Adding the contributions from the two categories, we find that the soft and collinear coefficients are
\begin{align}
c_{s,1}=-4C_F=-c_{c,1}\,,
\end{align}
and the final result for this term in the cross section is
%%%
\begin{equation}
\frac{1}{\sigma_0}\frac{\df\sigma^{(2,1)}}{\df\tau}
= 4C_F \ln\tau
\,,\end{equation}
%%%
which agrees with the well-known exact one-loop result for thrust~\cite{Ellis:1991qj}. This result was also reproduced in a somewhat different SCET framework in Ref.~\cite{Freedman:2013vya}, where the organization is different and there are more distinct contributions (17 terms) that add up to this leading logarithm result.

%~~~~~~~~~~~~~~~~~~~~~~~~~~~~~~~~~~~~~~~~~~~~~~~~~~~~~~~~~~~~~~~~~~~~~~~~~~~~~~~
\subsubsection{Power Corrections at NNLO}
%~~~~~~~~~~~~~~~~~~~~~~~~~~~~~~~~~~~~~~~~~~~~~~~~~~~~~~~~~~~~~~~~~~~~~~~~~~~~~~~

\begin{figure}[t]
\subfigure[]{\includegraphics[width=0.49\columnwidth]{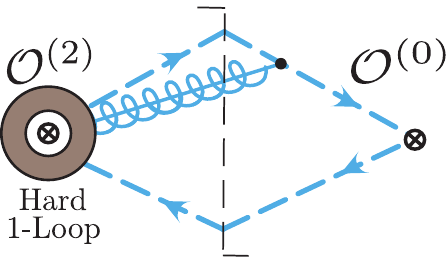}}
\hfill
\subfigure[]{\includegraphics[width=0.49\columnwidth]{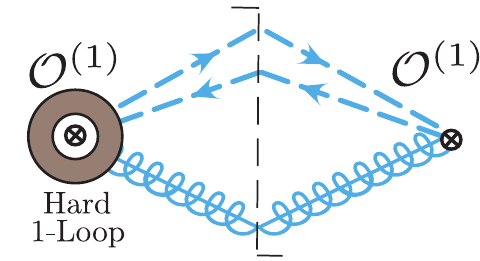}}
%%%
\caption{Representative diagrams of the two-loop hard collinear contributions which contribute at subleading power. Here the grey circle represents a one-loop hard virtual correction. There are contributions when either a gluon becomes collinear with a quark (a) or two quarks become collinear (b). The power suppression of the contributing operators is indicated.}
\label{fig:hc2loop}
\end{figure}

Having understood the structure at NLO, we can now extend our calculation to NNLO. The leading-logarithmic divergence at NNLO, $\alpha_s^2 \ln^3\tau$ can be obtained by dressing the NLO diagrams with a single leading-power correction that is either hard, collinear, or soft and could be virtual or real. From the consistency relations, summarized in \eq{constraints_final}, it suffices to calculate only the hard corrections to the collinear diagrams. This is shown schematically in \fig{hc2loop}. The loop corrections for $e^+e^-\to3$ partons are known analytically at both one~\cite{Ellis:1980wv} and two loops~\cite{Garland:2002ak}, which we use to derive our result. We find
%%%
\begin{align}
\frac{1}{\sigma_0}\frac{\df\sigma^{(2,2)}}{\df\tau}
&= \Bigl[ -32 C_F^2 +8C_F(C_F + C_A) \Bigr] \ln^3\tau
\nn \\
&=8 C_F (C_A-3 C_F) \ln^3\tau
\label{eq:thrustnnlo}
\,.\end{align}
%%%
In the first line we have separated the result into the contributions from Category 1, which have a $C_F^2$ color structure, as is the case for the leading logarithm at leading power, and the contribution from Category 2, which has a $C_F(C_F+C_A)$ color structure, which is not present at leading power. The appearance of this genuinely new color structure at subleading power is not surprising, due to the contribution from the limit where the two quarks become collinear, effectively giving rise to a $C_A$ like cusp, as seen in \fig{hc2loop}. It would be interesting to understand how this result can be derived from renormalization group evolution. 

We have performed an explicit calculation of the double soft coefficient $c_{ss,3}$ and the collinear-soft coefficient $c_{cs,3}$, and confirmed that the consistency relation in \eq{constraints_summary} indeed holds, providing a highly non-trivial check of our calculation. Knowing the double collinear coefficient $c_{cc,3}$ explicitly would further check the consistency relation, which we leave for future work.

%===============================================================================
\subsection{\boldmath $0$-jettiness in $q\bar q\to$ color-singlet production (beamthrust)}
\label{subsec:beam_thrust}
%===============================================================================

We now turn to computing the dominant subleading terms for $0$-jettiness or beam thrust. We define $q^\mu$, $Q$, and $Y$ as the total momentum, invariant mass, and rapidity of the color-singlet system,
%%%
\begin{equation}
Q = \sqrt{ q^2 }
\,,\qquad
Y = \frac{1}{2} \ln\frac{q^-}{q+}
\,.\end{equation}
%%%
The incoming partonic momenta are
%%%
\begin{align}
p_a &= x_a \Ecm \frac{n}{2}
\,,\qquad
x_a \Ecm = Q e^Y
\,, \nn \\
p_b &= x_b \Ecm \frac{\bn}{2}
\,,\qquad
x_b \Ecm = Q e^{-Y}
\,,\end{align}
%%%
where $n^\mu = (1, \hat z)$, $\bn^\mu = (1, -\hat z)$, and $\hat z$ is the beam axis.

It is important to distinguish different definitions of $0$-jettiness, which we refer to as leptonic and hadronic.
The dimensionful and dimensionless versions are defined as
%%%
\begin{align}
\Tau_0^x
&= \sum_k \min \Bigl\{ \lambda_x \,p_k^+, \lambda_x^{-1}\,p_k^- \Bigr\}
\,, \qquad
\tau^x \equiv \frac{\Tau_0^x}{Q}
\,,\end{align}
%%%
where the sum runs over all particles in the final state excluding the hard color-singlet system.
The momenta $p_k$ are defined in the hadronic center-of-mass frame and the measures are then defined as
%%%
\begin{align}\label{eq:Njet_def}
\text{leptonic:} && \lambda &= \sqrt{\frac{q^-}{q^+}} = e^Y
\,, \nn \\
\text{hadronic:} && \lambda_\hadcm &= 1
\,.\end{align}
%%%
For simplicity we will not include a superscript  on our default variable $\Tau_0\equiv \Tau_0^{\rm lept}$, which employs $\lambda = e^Y$.
The $e^Y$ factor in the leptonic definition is explicitly included to take into account the boost of the leptonic center-of-mass frame. It effectively defines $\Tau_0$ in the leptonic center-of-mass frame where $\hat p_k^\pm = e^{\pm Y} p_k^\pm$ and so $\Tau_0 = \sum_k\min\{\hat p_k^+, \hat p_k^-\}$. This is the more natural definition, as was discussed in detail already in Refs.~\cite{Stewart:2009yx, Stewart:2010tn, Stewart:2010pd}, and it is the definition used in \textsc{Geneva}~\cite{Alioli:2015toa} and the numerical results in Ref.~\cite{Gaunt:2015pea}.

The hadronic definition, which effectively defines $\Tau_0^\hadcm$ in the hadronic center-of-mass frame, was discussed in Refs.~\cite{Stewart:2009yx, Stewart:2010tn, Berger:2010xi} (called $\Tau_\mathrm{cm}$ there) for the purpose of experimental measurements where the total rapidity $Y$ is not known, e.g.\ due to the neutrinos in $W$ production or $H\to WW$. It is the definition currently used in MCFM8~\cite{Boughezal:2016wmq}.

In the context of $N$-jettiness subtractions, taking into account the boost $Y$ of the Born system is essential for ensuring that the power corrections are independent of $Y$. We therefore focus on the leptonic definition in this section. In \sec{discuss} we also discuss and compare to the hadronic definition, showing that it induces power corrections that are exponentially enhanced at large $Y$.

The hadronic cross section can be written as
%%%
\begin{equation}
\df \sigma
= \sum_{ij}\int\! \df \xi_a \df \xi_b\, f_i(\xi_a)\, f_j(\xi_b)\,\df\hat\sigma_{ij}(\xi_a,\xi_b)
\,,\end{equation}
%%%
where $f_i$ are the PDFs, and $\df\hat\sigma_{ij}(\xi_a,\xi_b)$ are the partonic cross sections.
We write the leading-order partonic cross section as
%%%
\begin{equation} \label{eq:sigmapartLO}
\frac{\df\hat\sigma_{q \bar q}^{(0,0)}(\xi_a, \xi_b; X)}{\df Q^2\, \df Y\,\df \tau}
= \sigma_{q0}(Q,X)\,\delta_a \delta_b \, \delta(\tau)
\,,\end{equation}
%%%
where we abbreviated
%%%
\begin{equation} \label{eq:deltaab}
\delta_a \equiv \delta(\xi_a - x_a)
\,, \qquad
\delta_b \equiv \delta(\xi_b - x_b)
\,.\end{equation}
%%%
In \eq{sigmapartLO}, $\sigma_{q0}(Q,X)$ is the Born cross section for the relevant $q\bar q \to $ color-singlet process mediated by the
$q\bar q $ vector current we consider.
It encodes the process dependence as well as all additional measurements and/or kinematic cuts $X$ applied to the color-singlet final state.

We note that for the NLP leading logarithms, the renormalization of the PDFs does not play a role, as they contain at most a single logarithm of $\tau$. This differs from the case of threshold resummation, where PDF renormalization contributes already to the leading logarithms.

A new feature at subleading power is that the PDF arguments develop sensitivity
to the small momentum components in the form
%%%
\begin{equation}
f_i\biggl[\xi\Bigl(1 + \frac{k}{Q}\Bigr)\biggr] = f_i(\xi) + \frac{k}{Q}\, \xi f_i'(\xi) + \dotsb
\,,\end{equation}
%%%
where $k/Q\sim \tau$. These are analogous to the phase space corrections for thrust discussed above, but arise due to the routing of a small momentum component through the incoming collinear lines. The PDF must thus be Taylor-expanded to achieve a homogeneous expansion in $\tau$. The first term corresponds to the leading-power contribution, while the second term contributes at next-to-leading power. It yields a $\ord{1}$ correction to the NLP coefficients even for small values of $\xi$, since $\xi f_i'(\xi) \sim f_i(\xi)$.

%~~~~~~~~~~~~~~~~~~~~~~~~~~~~~~~~~~~~~~~~~~~~~~~~~~~~~~~~~~~~~~~~~~~~~~~~~~~~~~~
\subsubsection{Results}
%~~~~~~~~~~~~~~~~~~~~~~~~~~~~~~~~~~~~~~~~~~~~~~~~~~~~~~~~~~~~~~~~~~~~~~~~~~~~~~~

We write the partonic cross section at subleading $\mathcal{O}(\tau)$ as
%%%
\begin{align} \label{eq:NLP}
\frac{\df\hat\sigma^{(2,n)}_{ij}(\xi_a, \xi_b; X)}{\df Q^2\,\df Y\,\df\tau}
= \sigma_{q0}(Q,X) \sum_{m=0}^{2n-1} C^{(2,n)}_{ij,m}(\xi_a, \xi_b)\ln^m\tau
\,,\end{align}
%%%
where $\sigma_{q0}(Q,X)$ is the Born cross section for the quark-initiated process, defined via \eq{sigmapartLO}.
For the coefficients we are interested in, there are in total six different partonic channels, $\hat\sigma_{q\bar q}, \hat\sigma_{\bar qq}, \hat\sigma_{qg}, \hat\sigma_{gq},\hat\sigma_{\bar qg},\hat\sigma_{g \bar q}$. They are all trivially related to the two basic channels $\hat\sigma_{q\bar q}$ and $\hat\sigma_{qg}$, on which we will focus.

To compute the leading-logarithmic coefficients $C^{(2,1)}_{ij,1}$ and $C^{(2,2)}_{ij,3}$, we cross our results for thrust computed in \subsec{thrust}. Taking into account the modified definition of the measurement function as well as the corrections from PDFs, we find for the NLO coefficients
%%%
\begin{align} \label{eq:NLOResult}
C_{q\bar q, 1}^{(2,1)}(\xi_a, \xi_b)
&=  8C_F\left(\delta_a \delta_b+\frac{\delta_a' \delta_b}{2}
+\frac{\delta_a \delta_b'}{2}\right)
\,, \\
C_{qg, 1}^{(2,1)}(\xi_a, \xi_b)
&= -2 T_F \,\delta_a \delta_b
\,,\end{align}
%%%
where $T_F=1/2$. The $qg$ channel has a different color factor due to differences in the averages over initial state
colors. The derivatives of the delta functions are defined as
%%%
\begin{align}
\delta'_a \equiv x_a\, \delta'(\xi_a - x_a)
\,, \qquad
\delta'_b \equiv x_b\, \delta'(\xi_b - x_b)
\,,\end{align}
%%%
which translate into the above-mentioned PDF derivatives in the hadronic cross section.
They only appear in the $q\bar q$ coefficients, because the $qg$ coefficient has no analog at leading power that is sufficiently singular.

\begin{figure}[t]
\subfigure[]{\includegraphics[width=0.45\columnwidth]{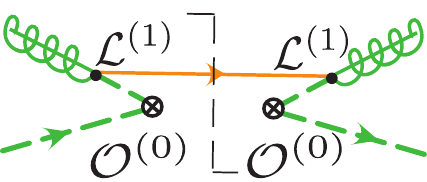}}
\hfill
\subfigure[]{\includegraphics[width=0.45\columnwidth]{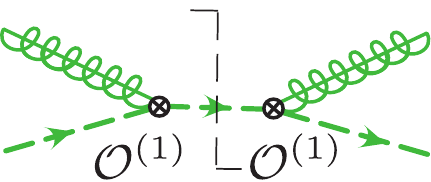}}
%%%
\caption{Diagrams contributing to $\sigma_{qg}$, where either a soft quark crosses the cut (a), or a collinear quark crosses the cut (b). The one-loop corrections to these diagrams give rise to the $T_F(C_F+C_A) \ln^3(\tau)$ correction to beam thrust.}
\label{fig:beamthrust}
\end{figure}

Repeating the analysis at NNLO, we obtain for the leading-logarithmic coefficients
%%%
\begin{align} \label{eq:NNLOResult}
C_{q\bar q, 3}^{(2,2)}(\xi_a, \xi_b)
&= -32 C_F^2 \left(\delta_a \delta_b+\frac{\delta_a' \delta_b}{2}
+\frac{\delta_a \delta_b'}{2}\right)
\,, \\
C_{qg, 3}^{(2,2)}(\xi_a, \xi_b)
&=  4 T_F(C_F+C_A)\, \delta_a \delta_b
\,, \end{align}
%%%
which are one of the main results of this paper. The result for the $q\bar q$ channel has a $C_F^2$ color structure, which is analogous to the leading logarithm at leading power. On the other hand, the $qg$ channel has a $T_F(C_F+C_A)$ color structure. This arises from the one-loop corrections to the diagrams in \fig{beamthrust}, where a soft or collinear quark crosses the cut.

We have also analytically calculated the $\alpha_s^2 \ln^3(1-z)$ corrections to Drell-Yan production in the threshold limit for both the $q\bar q$ and $qg$ channels, for which we find agreement with the known NNLO results \cite{Hamberg:1990np} for all color structures in both channels. This provides a highly nontrivial cross check of our approach. While subleading power corrections in the threshold limit have been well studied for the $q\bar q$ channel \cite{Laenen:2008ux, Laenen:2010uz, Bonocore:2014wua, Bonocore:2015esa, Bonocore:2016awd}, we are not aware of studies for the $qg$ channel.

Further details of the SCET-based NNLO analysis discussed here  for both $0$-jettiness and the threshold limit will be provided in a future dedicated publication.

\begin{figure*}[t!]
\includegraphics[width=\columnwidth]{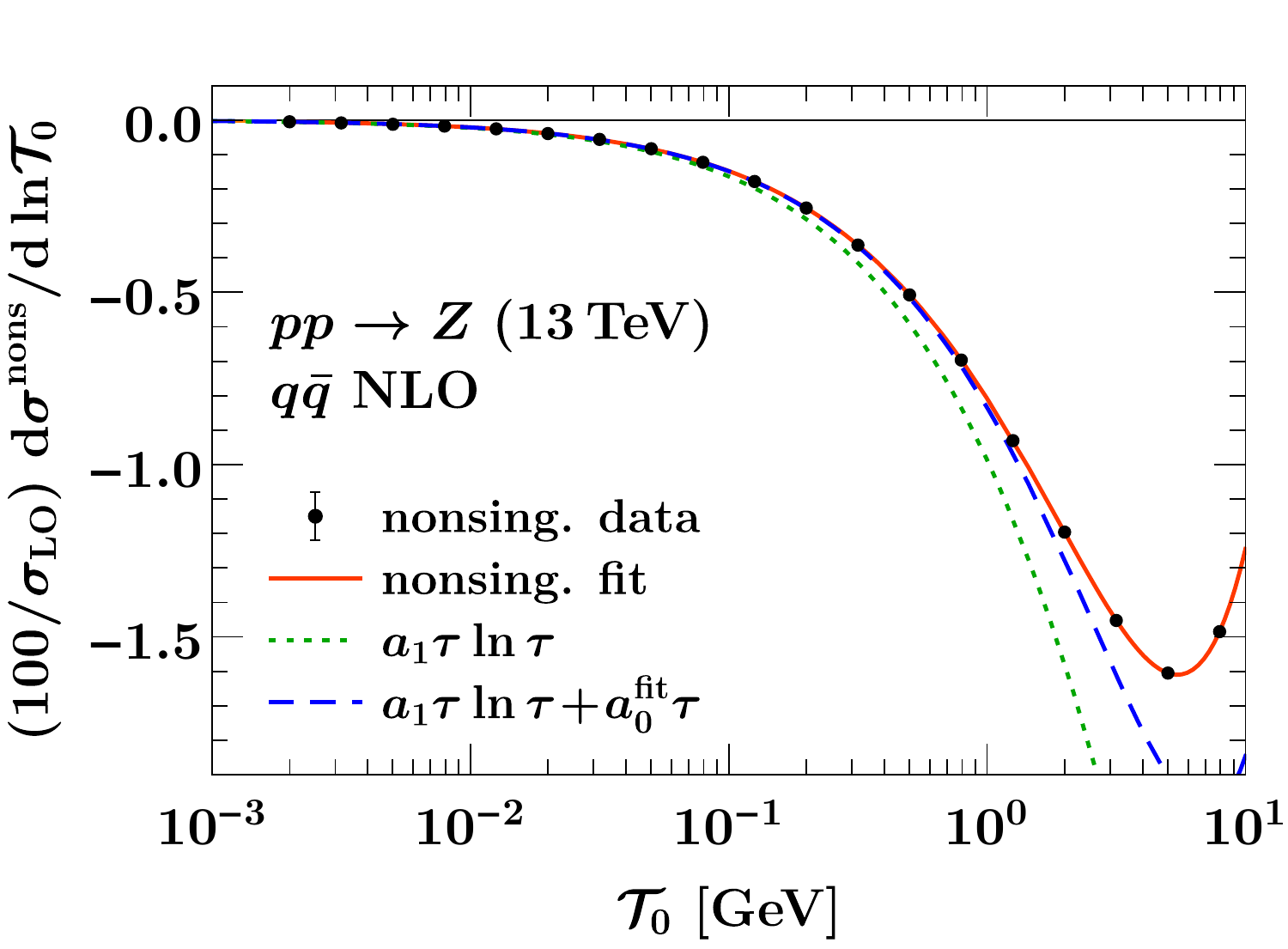}%
\hfill
\includegraphics[width=\columnwidth]{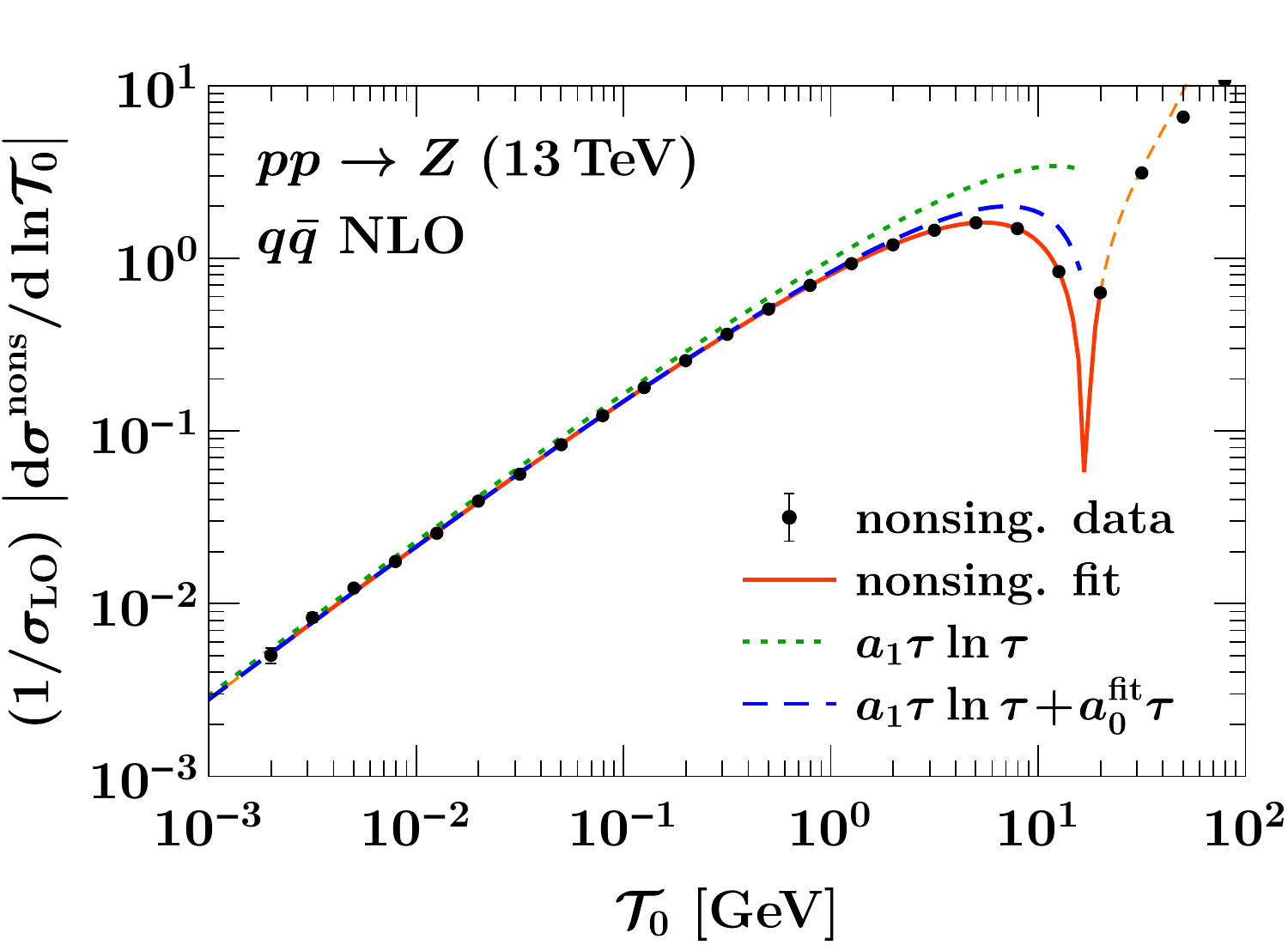}%
\\
\includegraphics[width=\columnwidth]{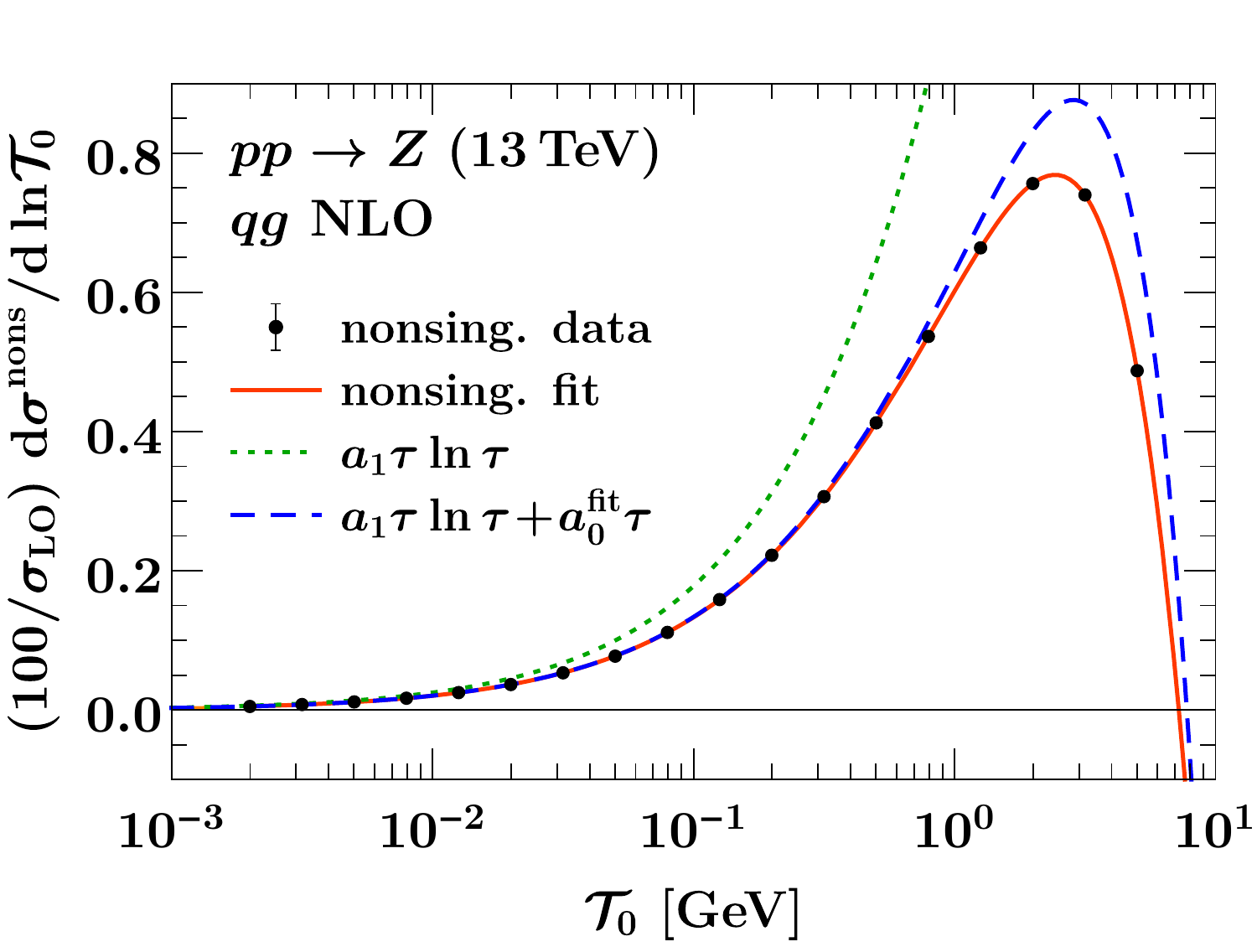}%
\hfill
\includegraphics[width=\columnwidth]{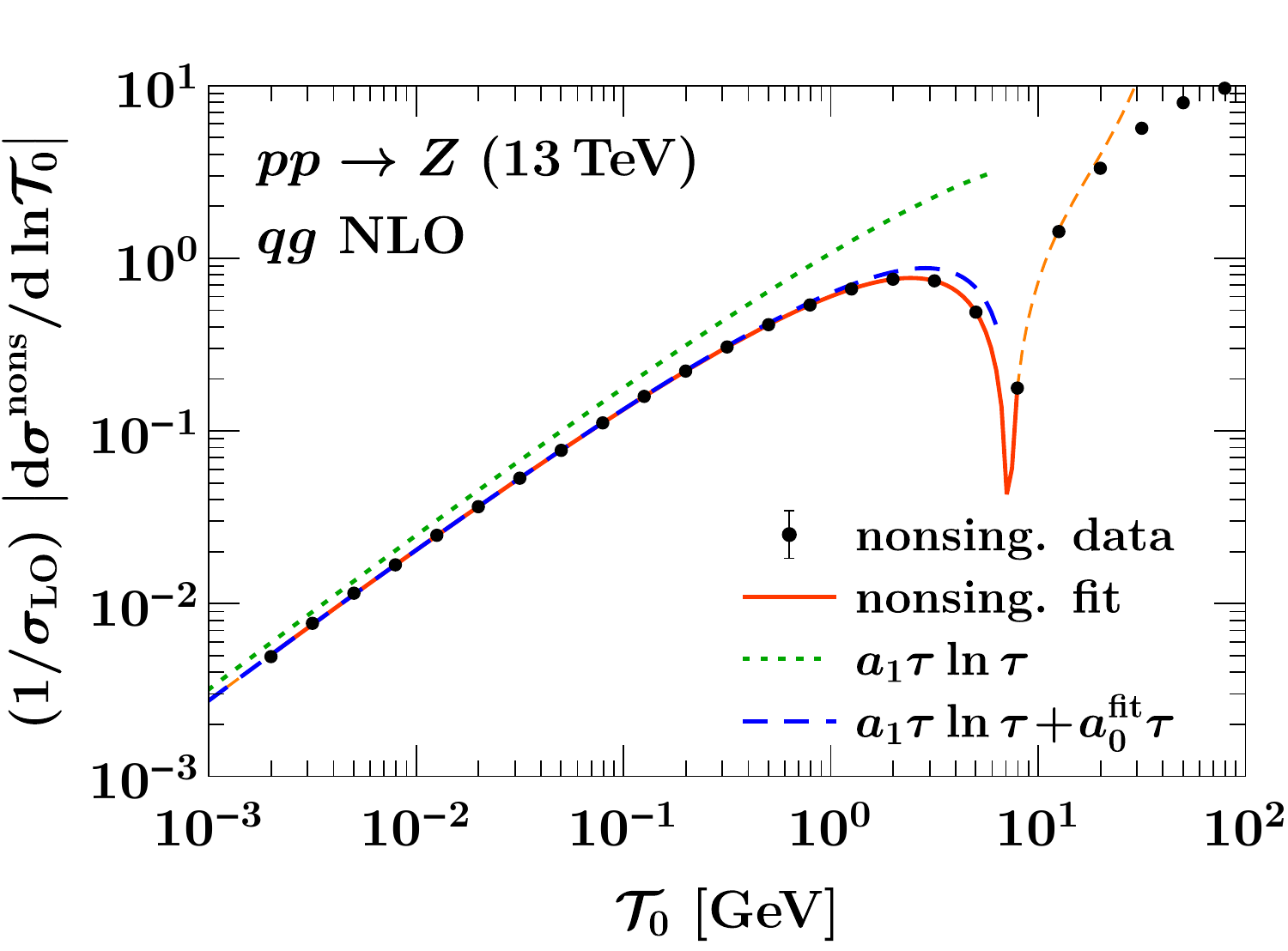}%
%%%
\caption{Illustration of the fit to the $\ord{\alpha_s}$ nonsingular in the $q\bar q$ channel (top row) and the $qg$ channel (bottom row). The plots on the right are equivalent to those on the left and show the absolute value on a logarithmic scale. A detailed explanation of the fit function, as well as the plotted curves, is given in the text.}
\label{fig:fitNLO}
\end{figure*}

\begin{figure*}[t!]
\includegraphics[width=\columnwidth]{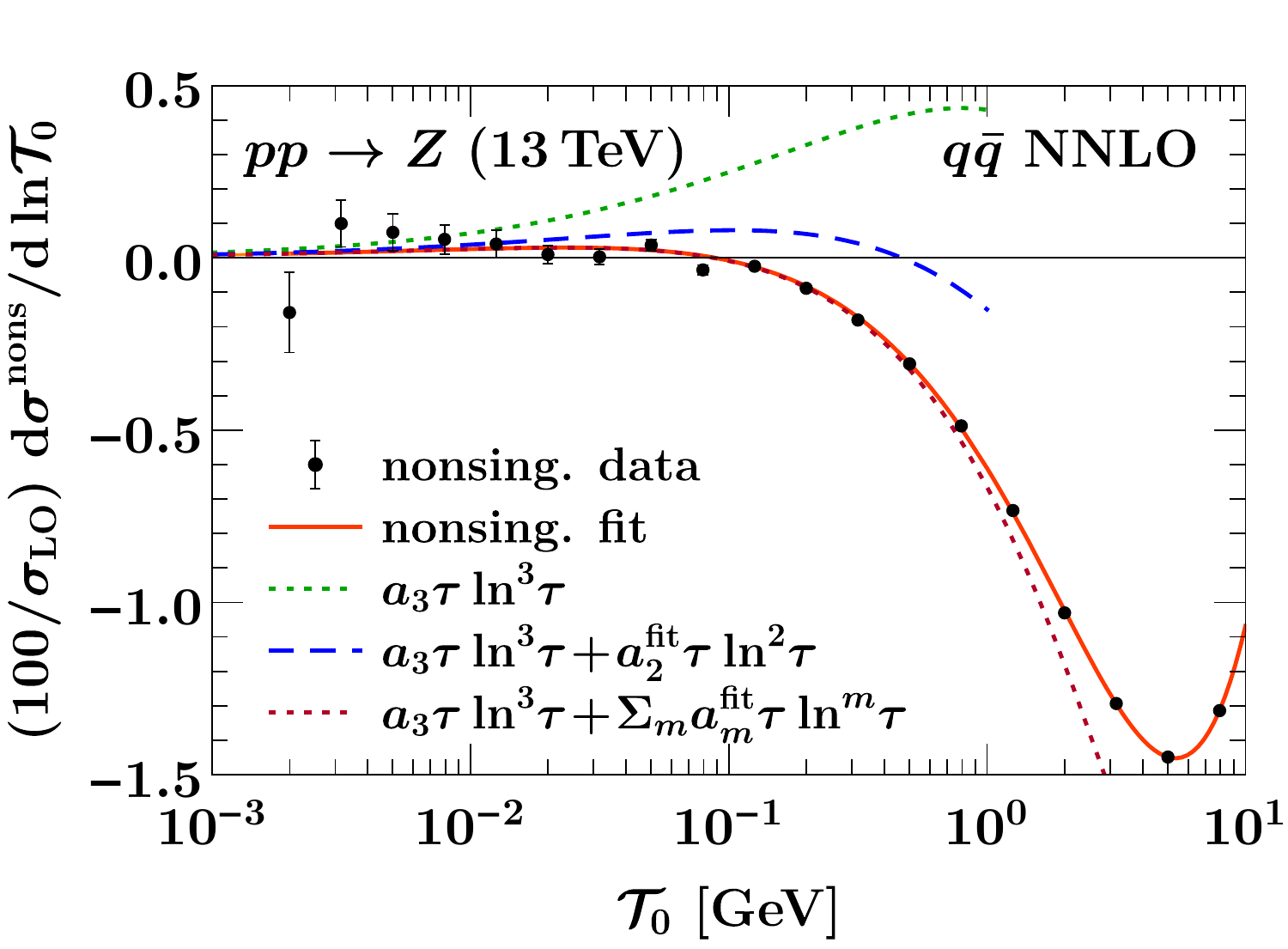}%
\hfill
\includegraphics[width=\columnwidth]{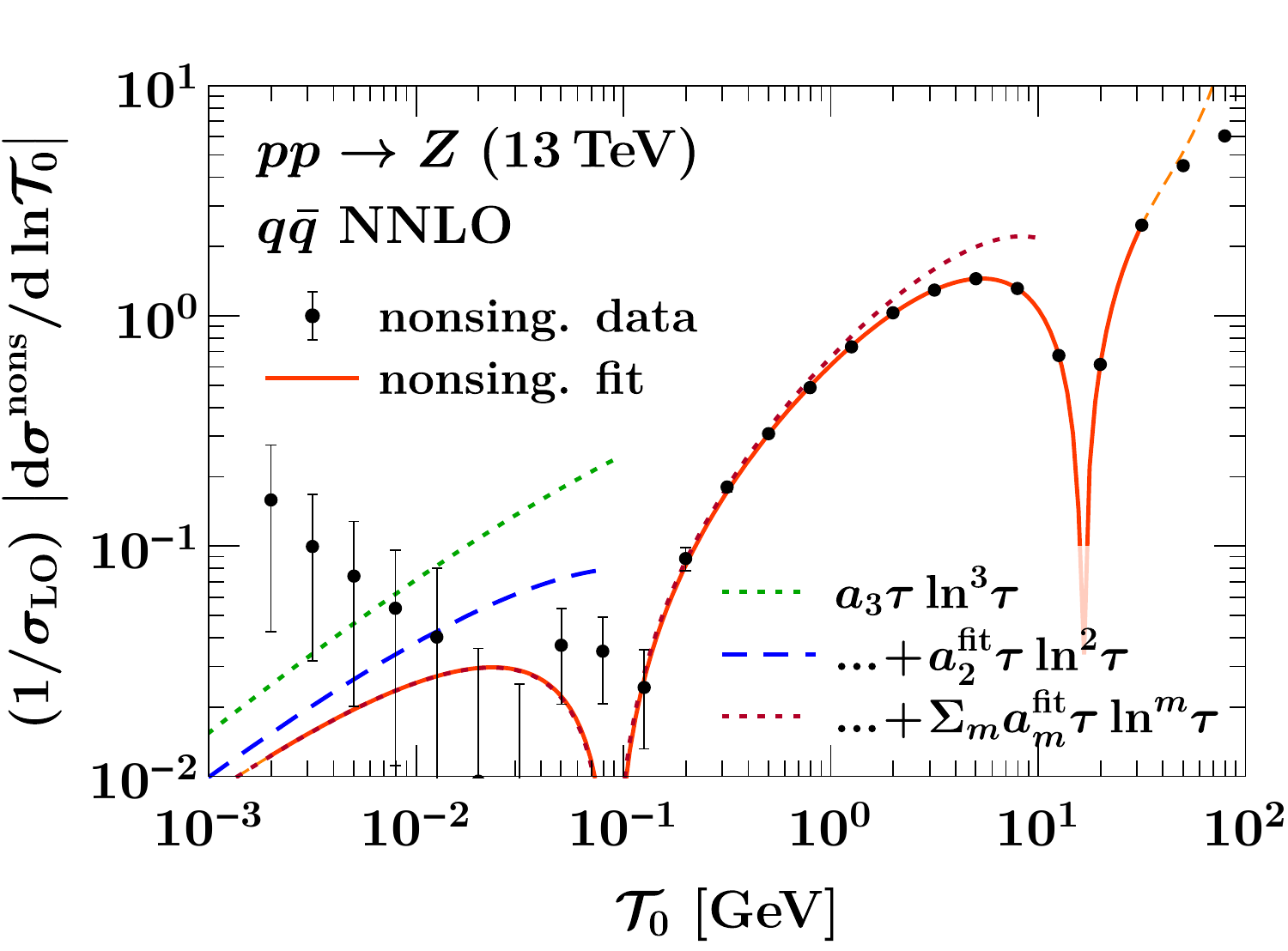}%
\\
\includegraphics[width=\columnwidth]{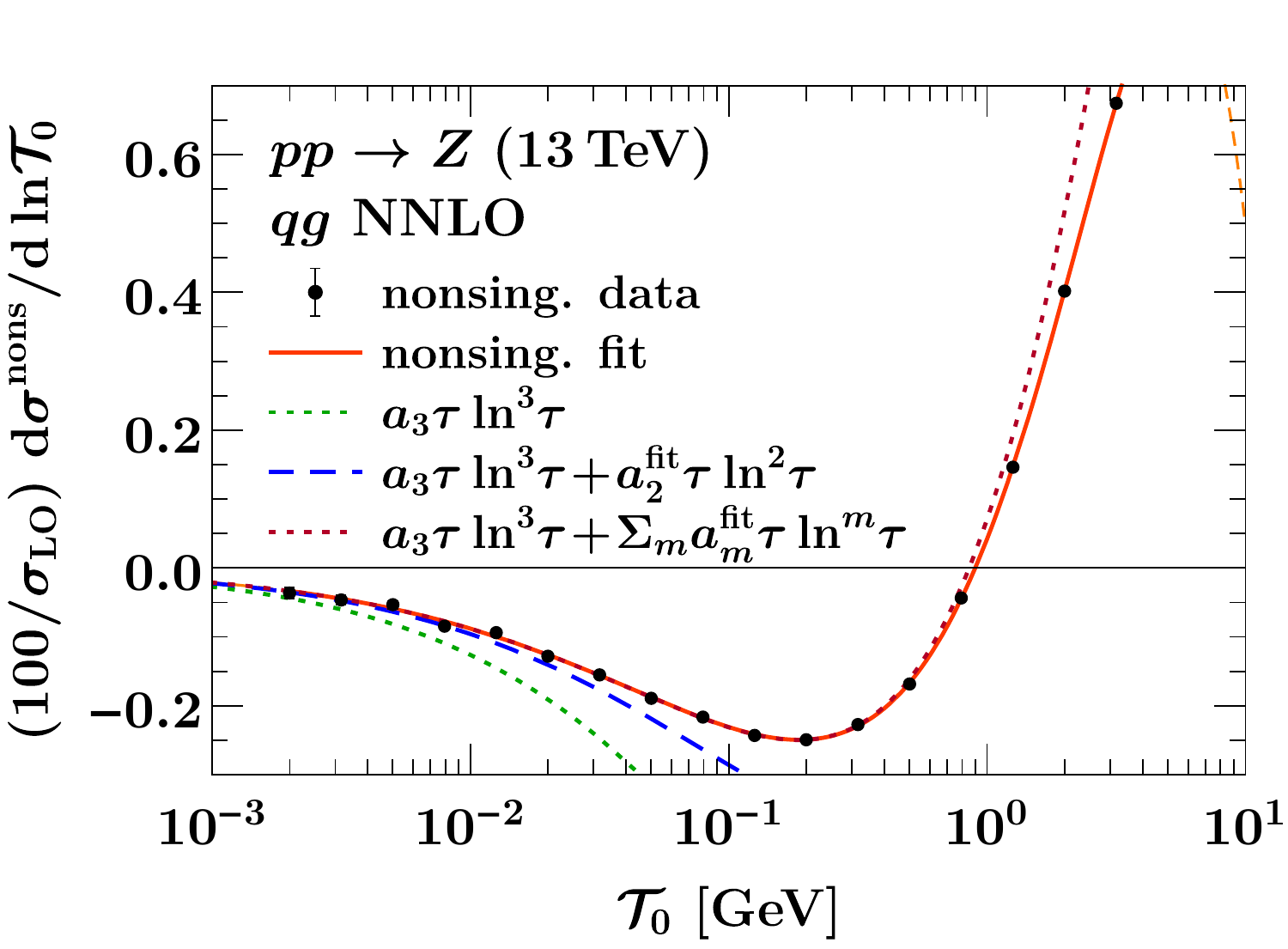}%
\hfill
\includegraphics[width=\columnwidth]{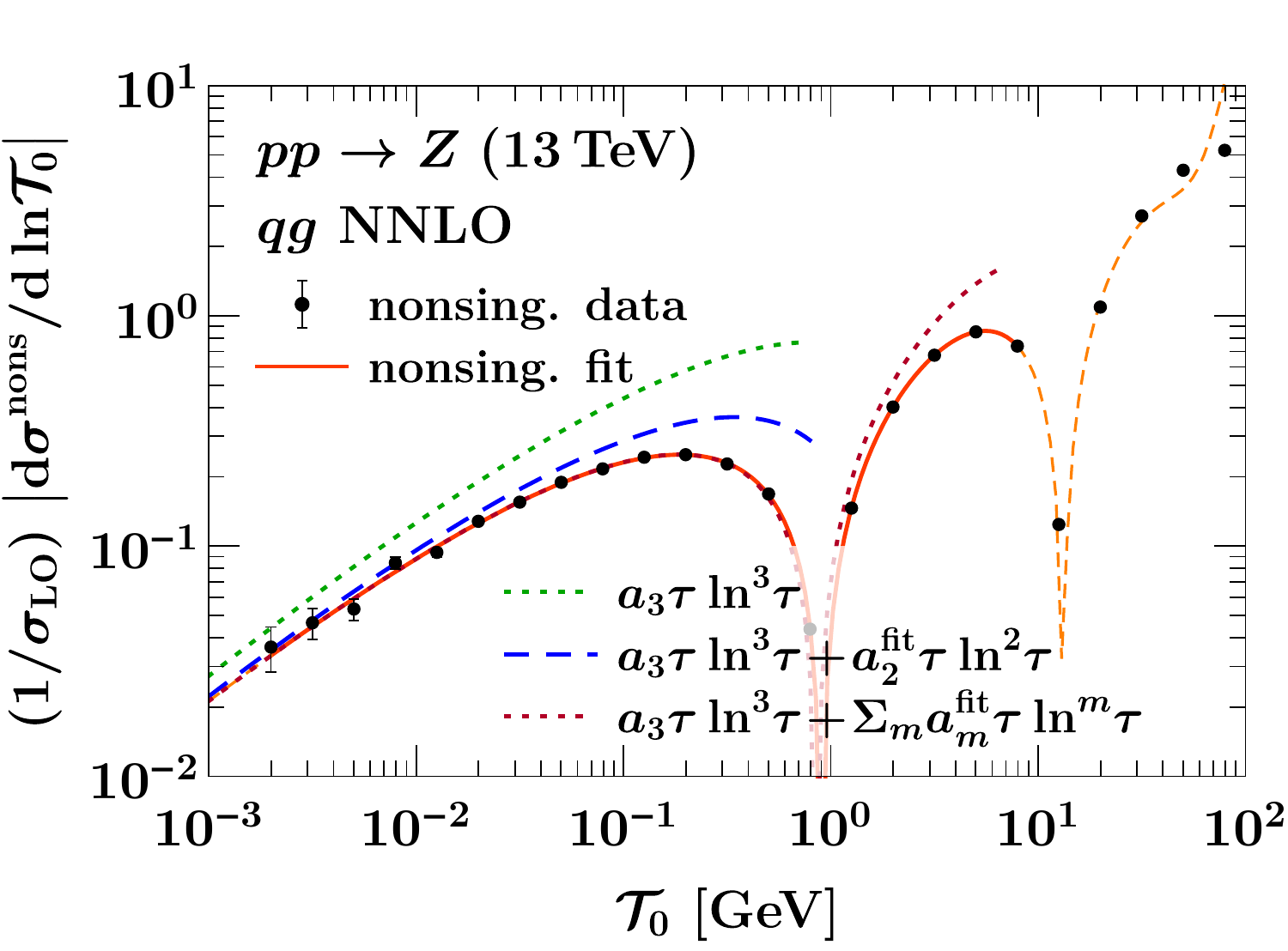}%
%%%
\caption{Illustration of the fit to the $\ord{\alpha_s^2}$ nonsingular in the $q\bar q$ channel (top row) and the $qg$ channel (bottom row). The plots on the right are equivalent to those on the left and show the absolute value on a logarithmic scale. A detailed explanation of the fit function, as well as the plotted curves, is given in the text.}
\label{fig:fitNNLO}
\end{figure*}

\begin{figure*}[t]
\includegraphics[width=\columnwidth]{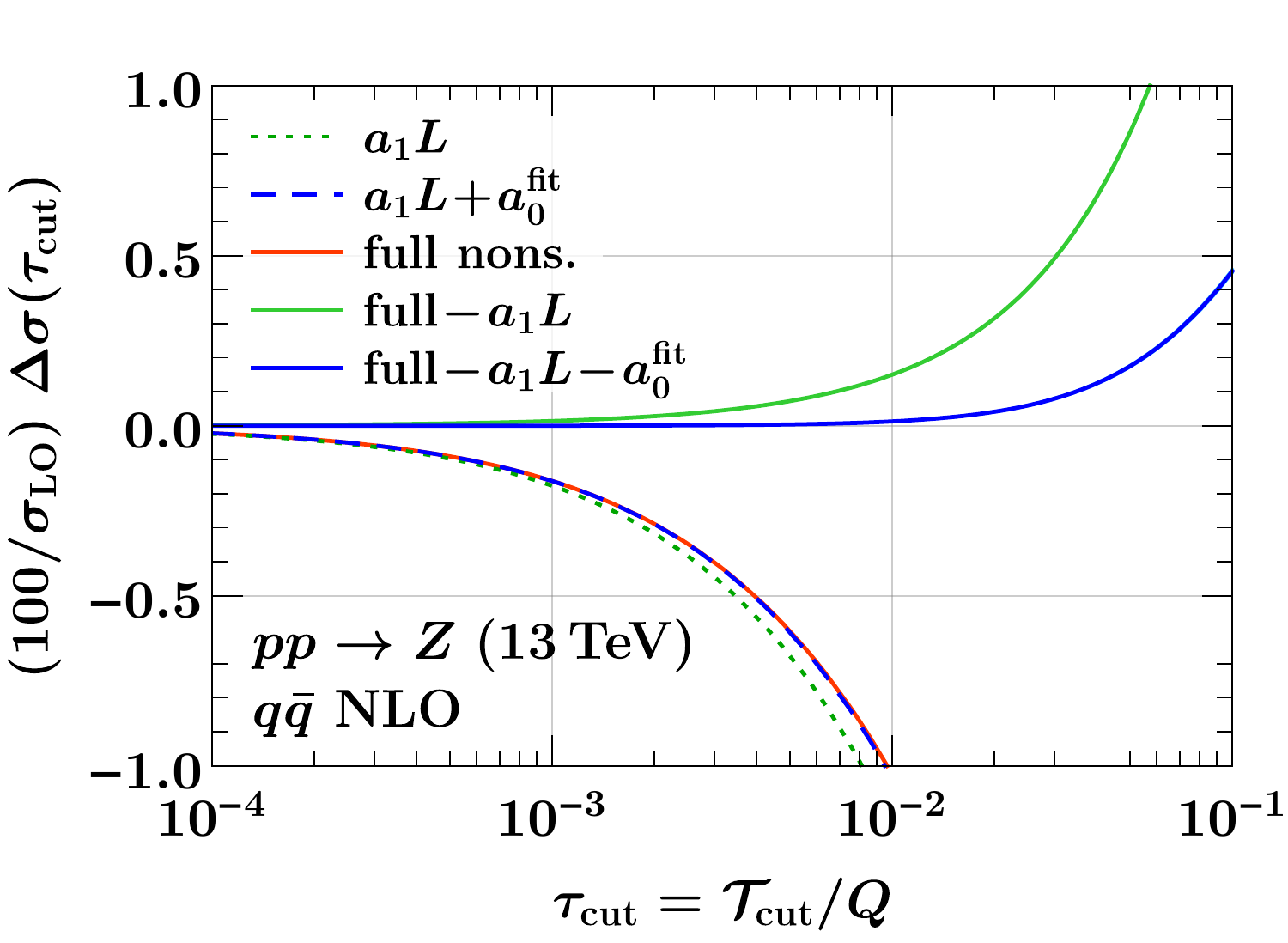}%
\hfill
\includegraphics[width=\columnwidth]{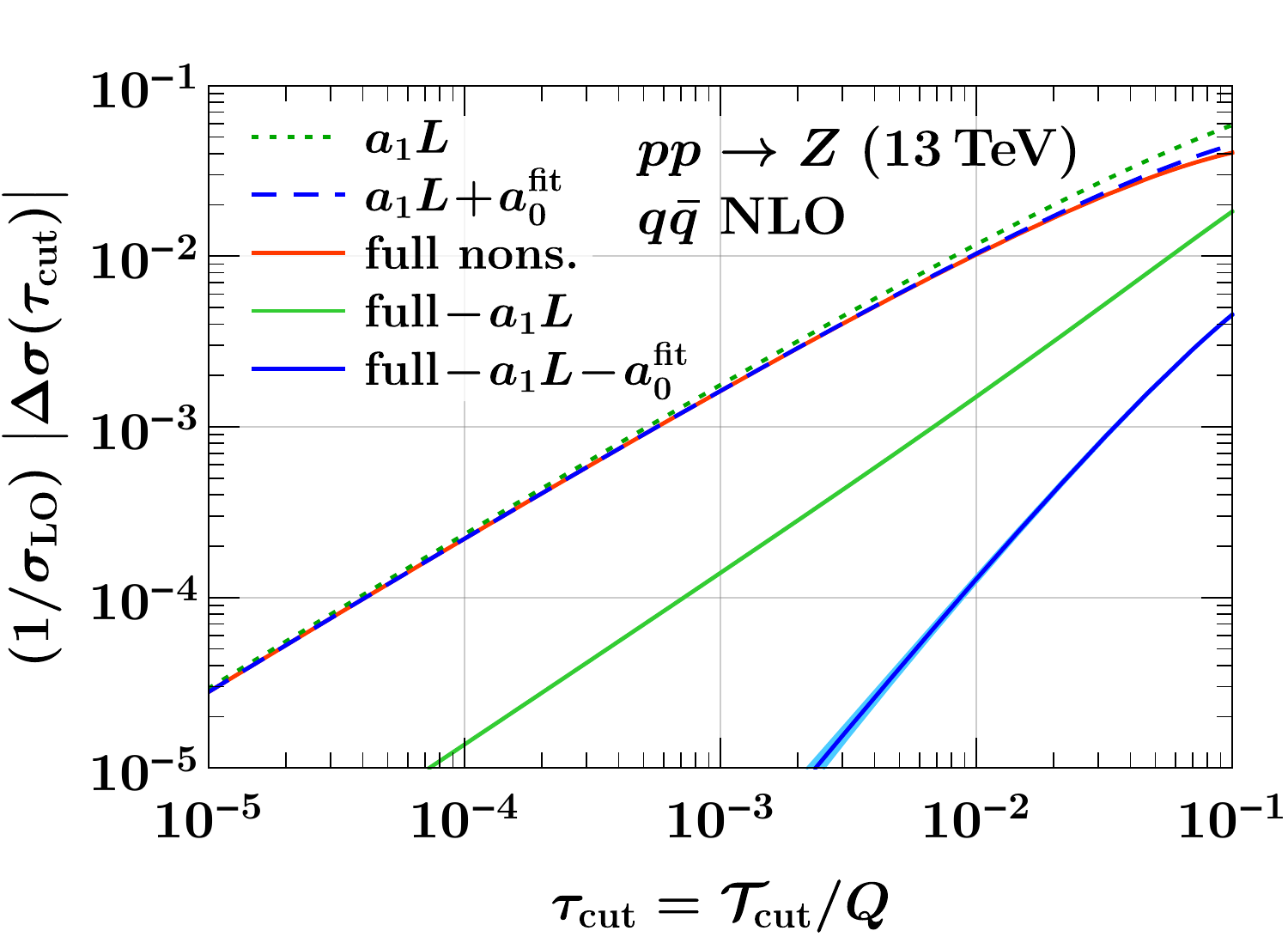}%
\\
\includegraphics[width=\columnwidth]{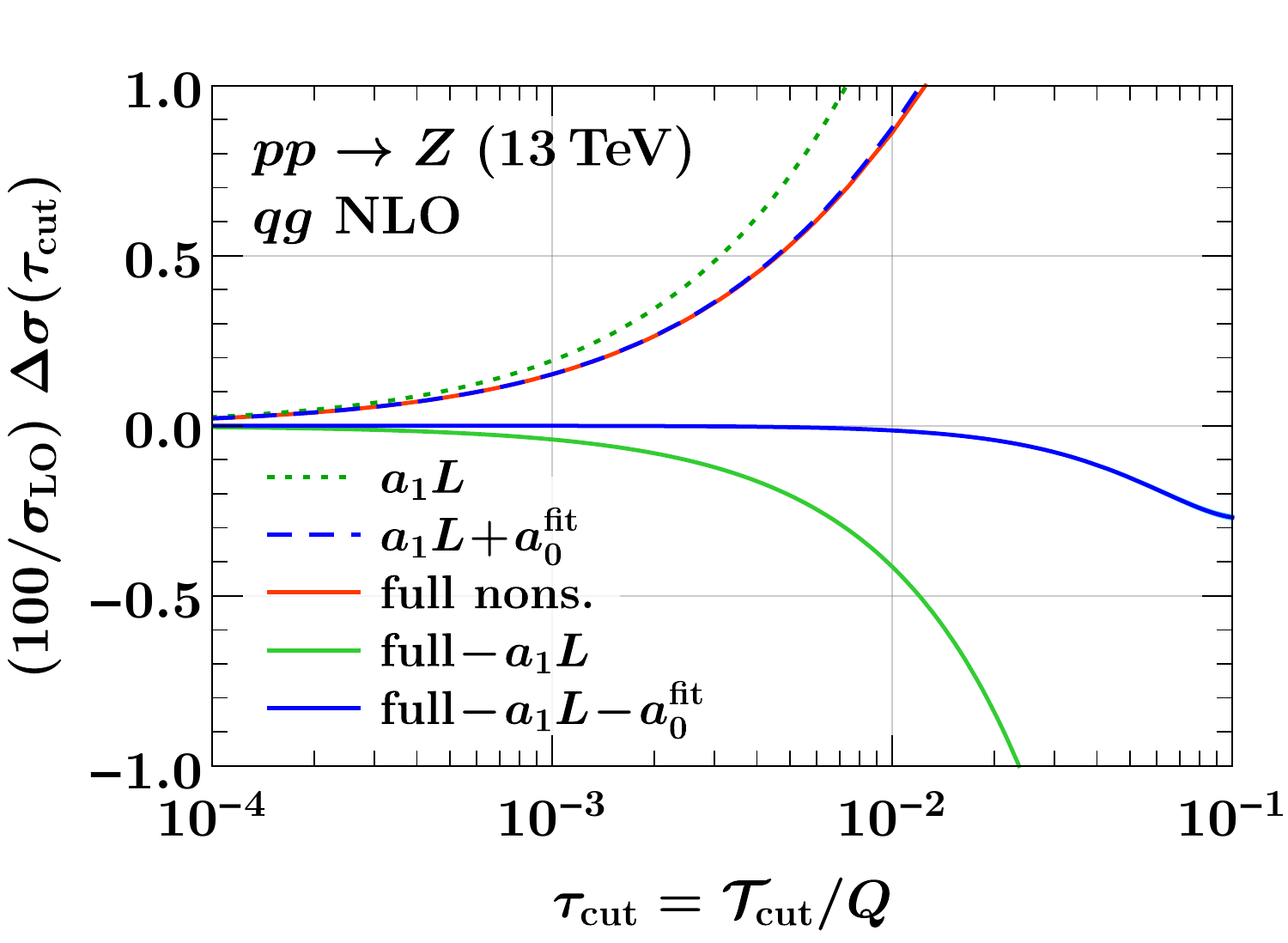}%
\hfill
\includegraphics[width=\columnwidth]{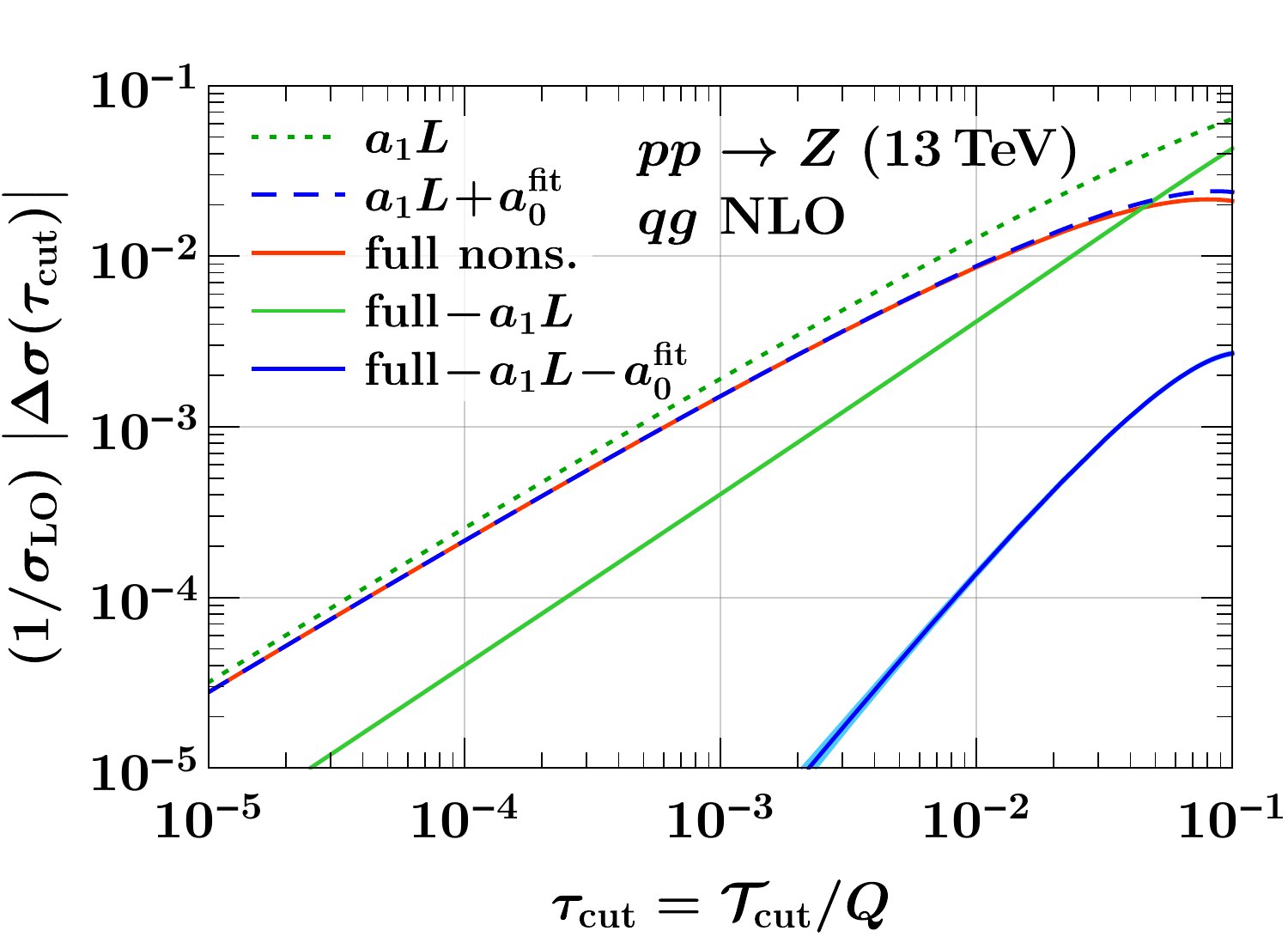}%
%%%
\caption{Power corrections $\Delta\sigma(\tau_\cut)$ for the $\ord{\alpha_s}$ contributions in the $q\bar q$ channel (top row) and the $qg$ channel (bottom row). The plots on the right are equivalent to those on the left and show the absolute value on a logarithmic scale.}
\label{fig:cumulantNLO}
\end{figure*}

\begin{figure*}[t]
\includegraphics[width=\columnwidth]{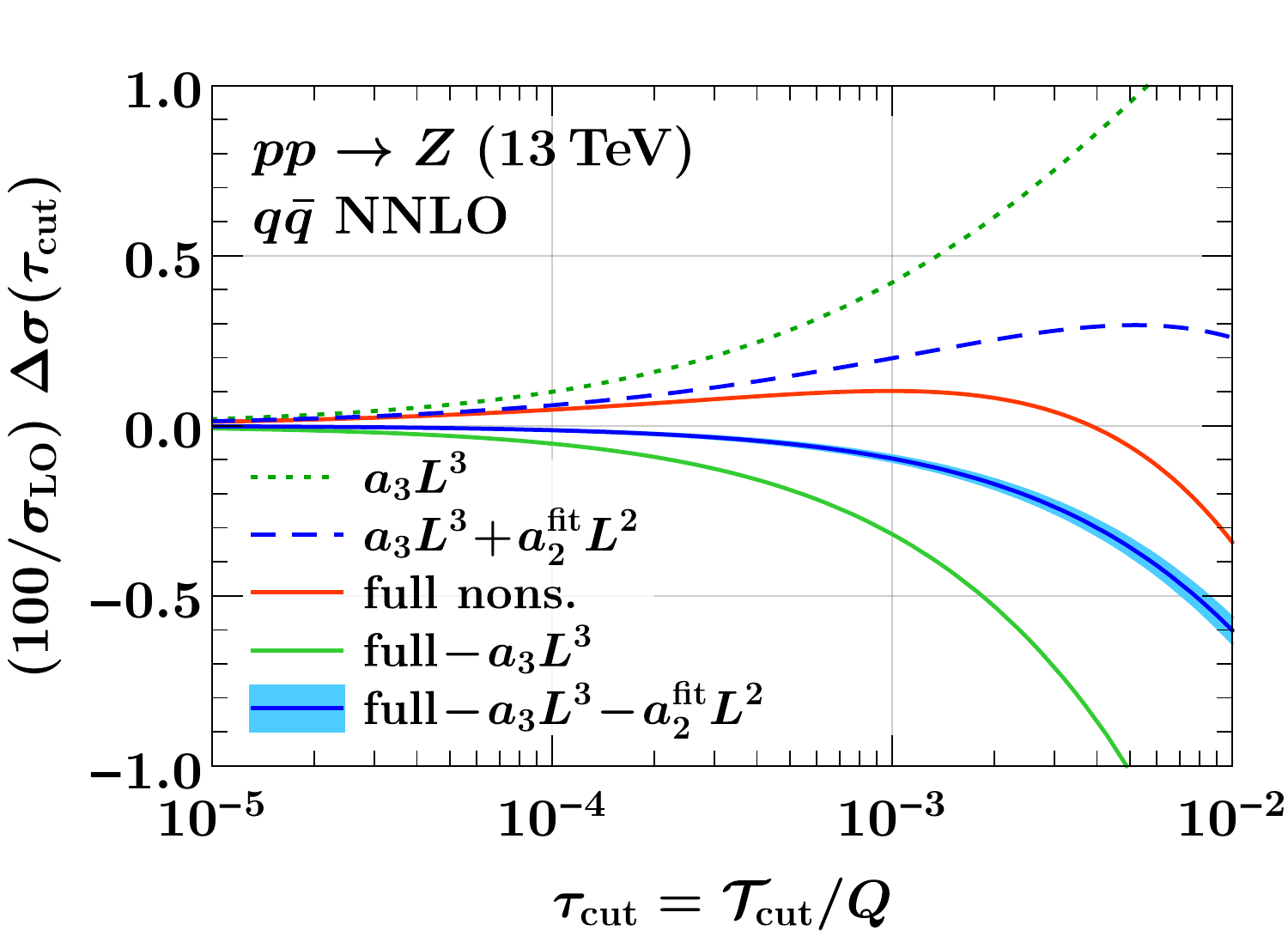}%
\hfill
\includegraphics[width=\columnwidth]{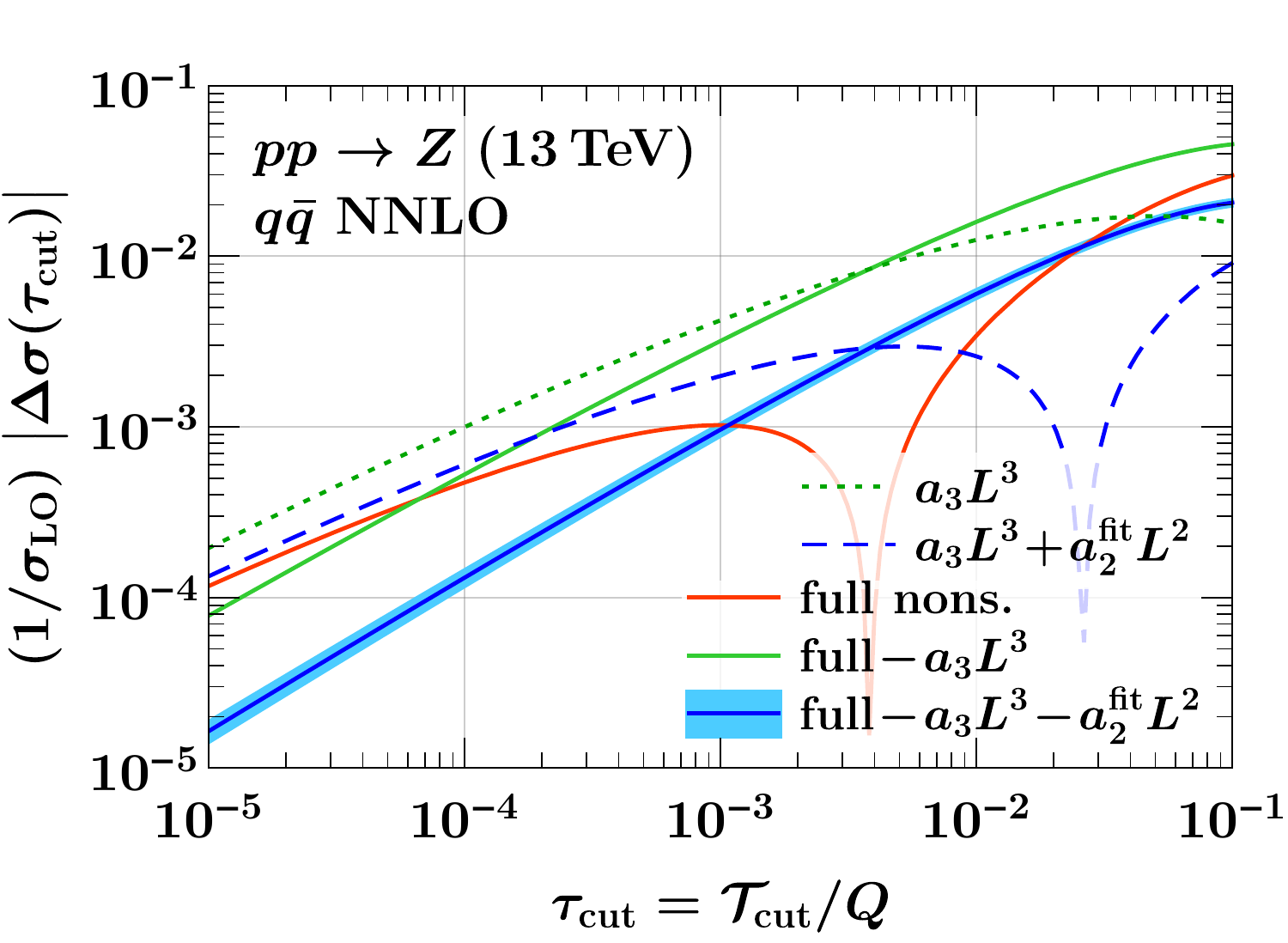}%
\\
\includegraphics[width=\columnwidth]{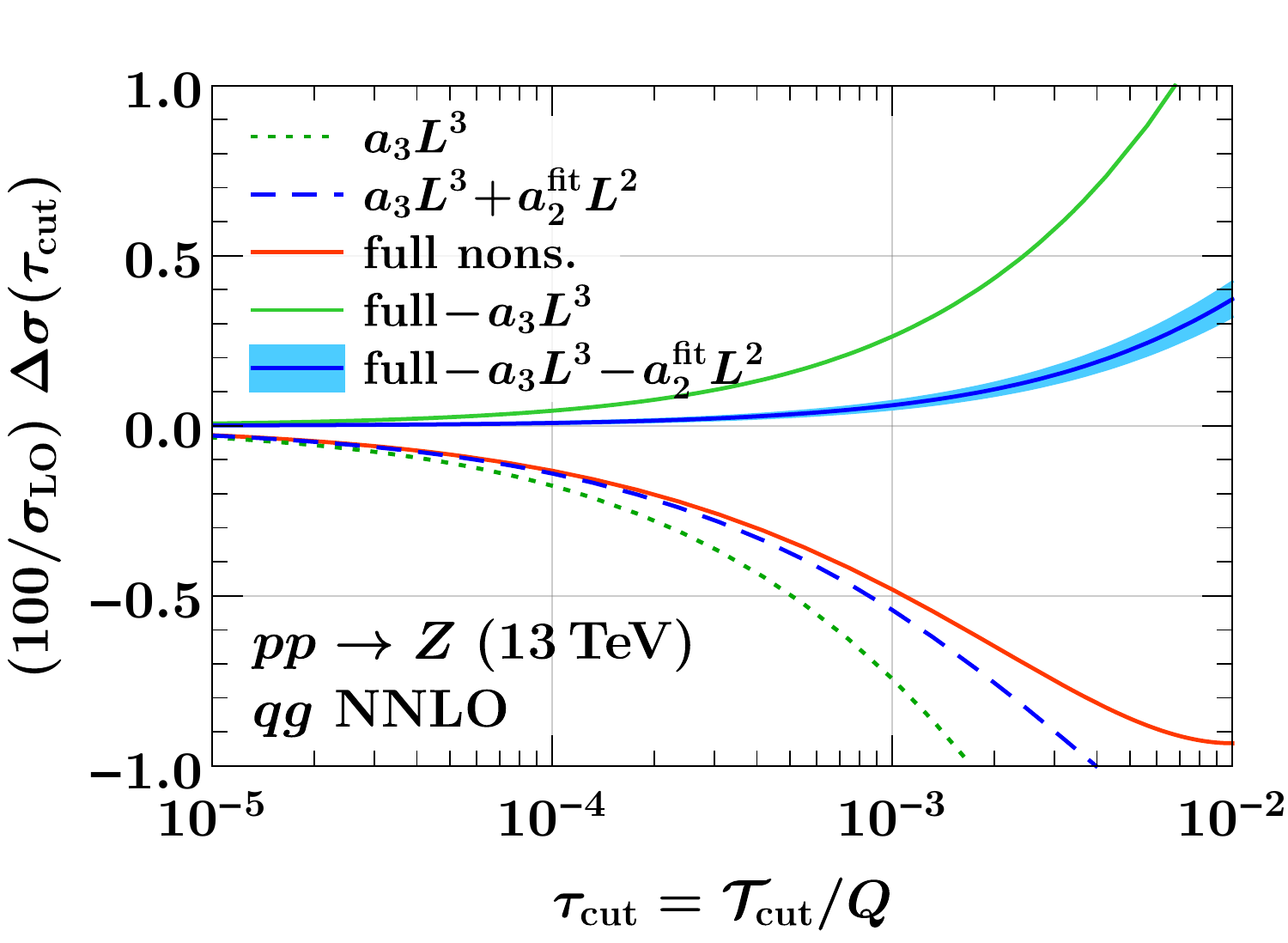}%
\hfill
\includegraphics[width=\columnwidth]{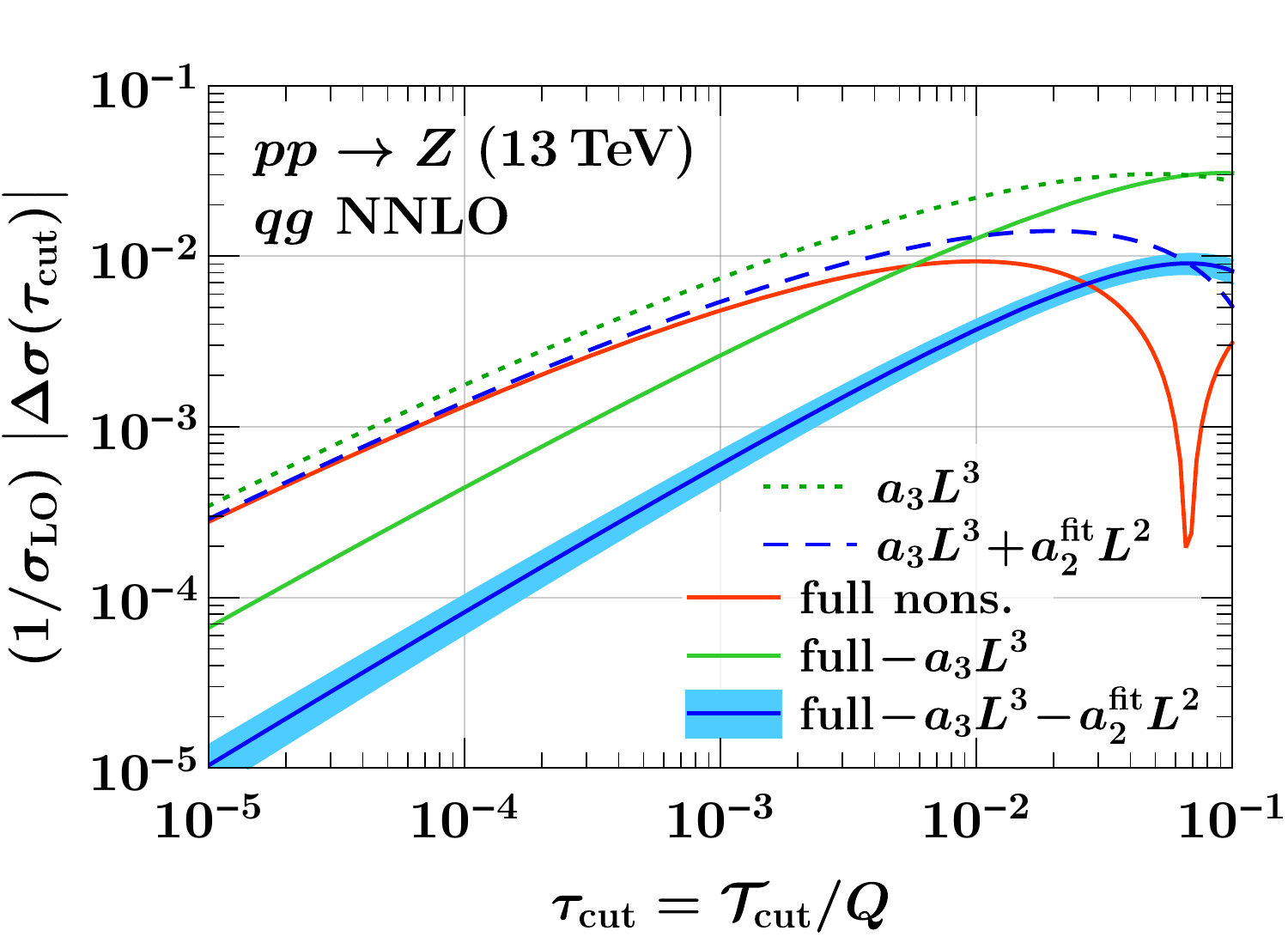}%
\\
\includegraphics[width=\columnwidth]{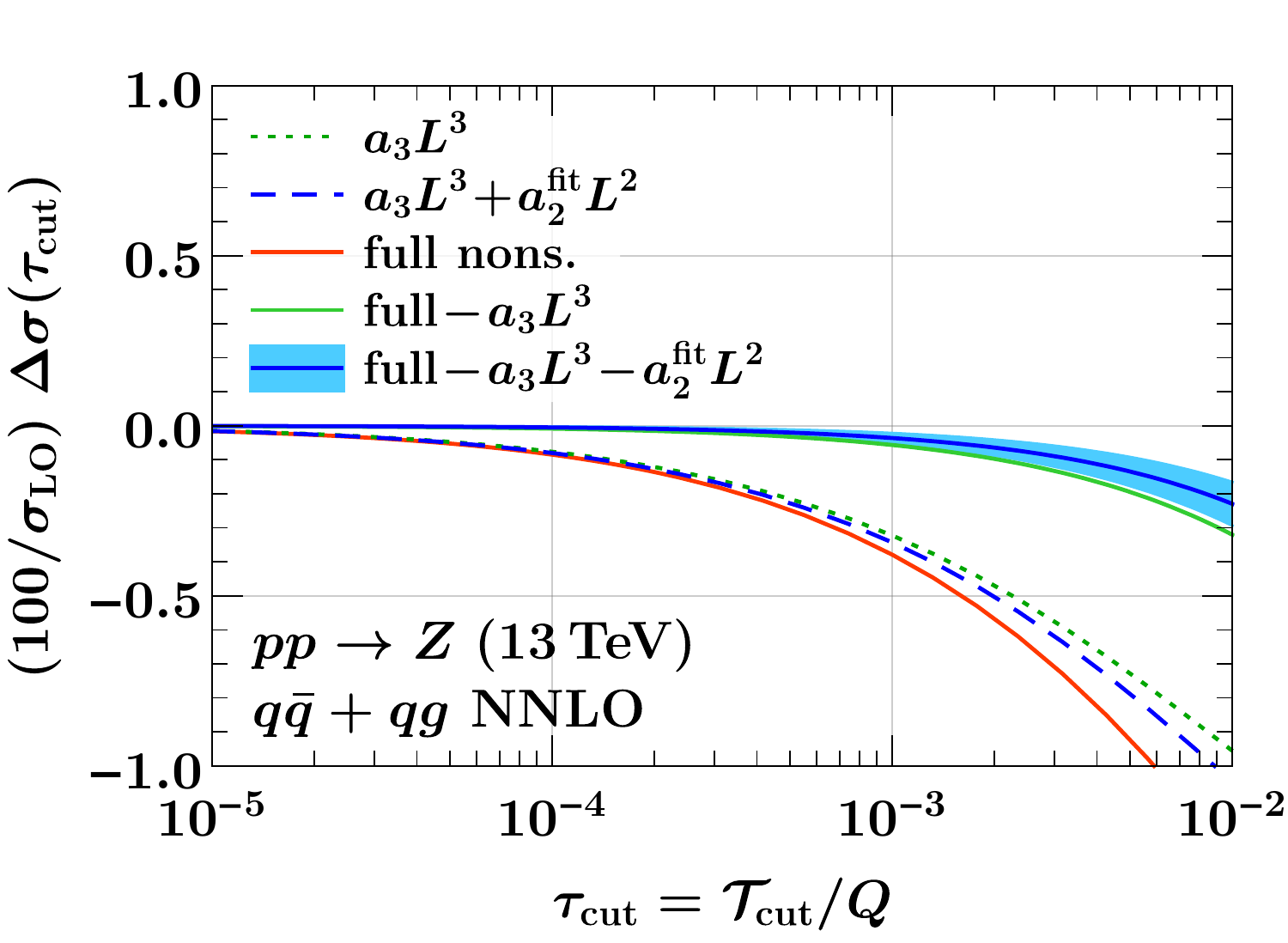}%
\hfill
\includegraphics[width=\columnwidth]{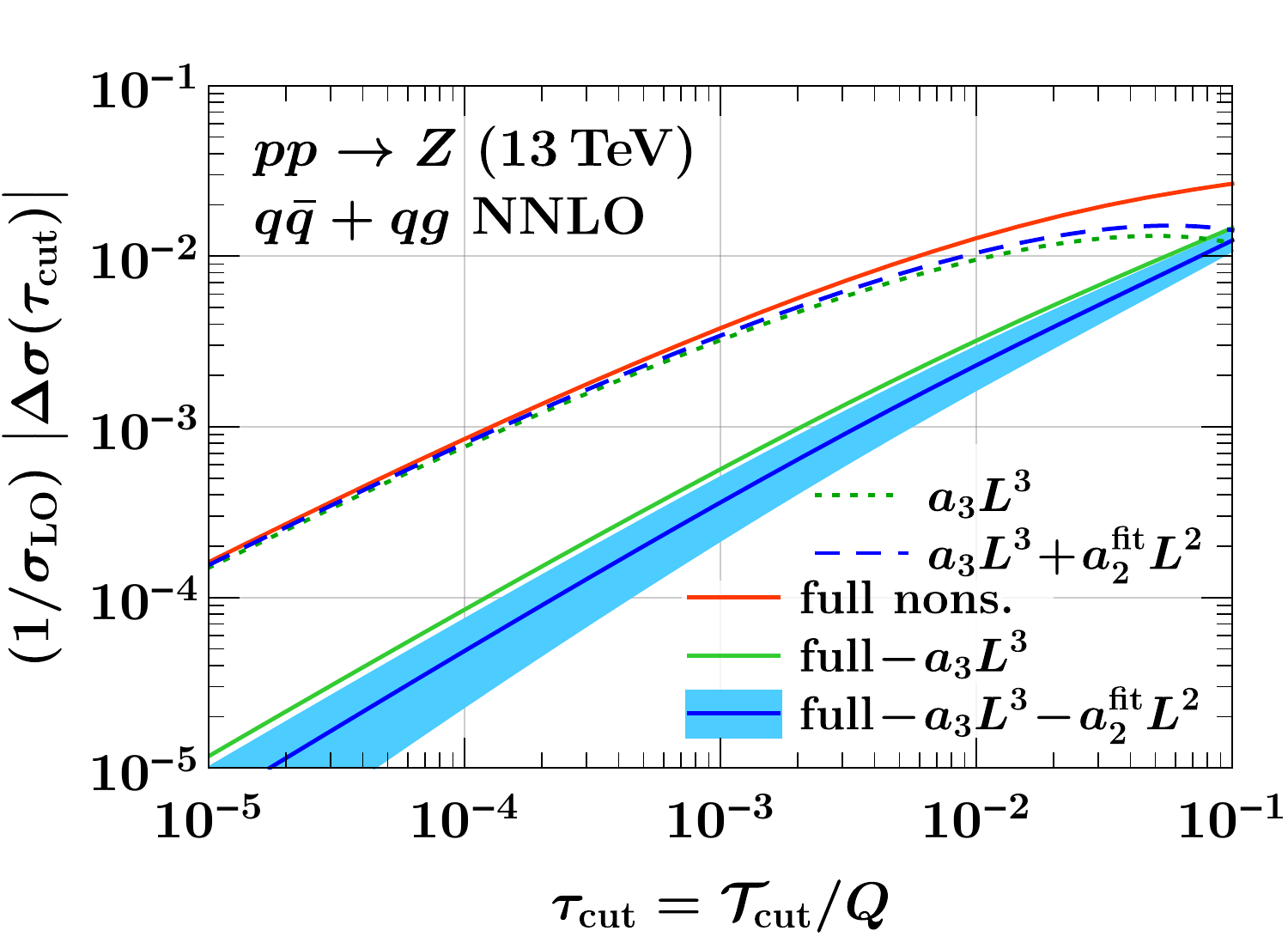}%
%%%
\caption{Power corrections $\Delta\sigma(\tau_\cut)$ for the $\ord{\alpha_s^2}$ contributions in the $q\bar q$ channel (top row), the $qg$ channel (middle row), and the sum of both channels (bottom row).
The plots on the right are equivalent to those on the left and show the absolute value on a logarithmic scale.}
\label{fig:cumulantNNLO}
\end{figure*}

%~~~~~~~~~~~~~~~~~~~~~~~~~~~~~~~~~~~~~~~~~~~~~~~~~~~~~~~~~~~~~~~~~~~~~~~~~~~~~~~
\subsubsection{Numerical results}
%~~~~~~~~~~~~~~~~~~~~~~~~~~~~~~~~~~~~~~~~~~~~~~~~~~~~~~~~~~~~~~~~~~~~~~~~~~~~~~~

We now present a comparison of our results with numerical fixed-order results.
On the one hand, this serves as a numerical cross check on our
calculated coefficients for the leading logarithm at subleading power. On the other hand, using our exact
results for the calculated coefficients allows us to numerically extract the NLL corrections
of $\ord{\alpha_s \tau}$ and $\ord{\alpha_s^2\tau \ln^2\tau}$, and to study the numerical relevance
of both the LL and NLL terms at subleading power.

For our numerical analysis we consider the process $pp\to Z/\gamma^*$ 
at $E_{\rm cm} = 13\TeV$. We always use the MMHT2014 NNLO PDFs~\cite{Harland-Lang:2014zoa},
fixed scales $\mu_r = \mu_f = m_Z$, and $\alpha_s(m_Z) = 0.118$.
We work at fixed $Q = m_Z$ in the narrow-width approximation,
which avoids the numerical phase-space integral over $Q$. We integrate over the full range of
rapidity $\ln(m_Z/E_{\rm cm}) \leq Y \leq -\ln(m_Z/E_{\rm cm})$ and are fully inclusive over
the vector-boson decay. (We include the branching ratio to leptons in $\sigma_{q0}$, but this only affects the
overall cross section and is irrelevant for our studies.)
The leading-order cross section following from \eq{sigmapartLO} is then given by
%%%
\begin{equation}
\sigma_\LO
= \pi\Gamma_Z m_Z \sum_q \sigma_{q0}(m_Z)\int\!\df Y f_q\bigl(m_Z e^Y\bigr) f_{\bar q}\bigl(m_Z e^{-Y}\bigr)
\,,\end{equation}
%%%
where the sum over $q$ runs over all quark flavors $q=\{d,u,s,c,b, \bar d, \bar u, \bar s, \bar c, \bar b\}$.

We normalize all our results to $\sigma_\LO$, which removes a large part of the dependence on the explicit process that is mediated by the underlying $q \bar q$ vector current.
The only leftover dependence is related to the PDFs and comes from the effective $x$-range in the PDFs probed by the rapidity integration, which is determined by the value of $Q = m_Z$, as well as the included quark flavors. These PDF effects primarily determine the size of the $q\bar q$ and $qg$ channels relative to each other, but only to a small extent the size of the corrections within a channel. (The PDF derivative contributions in the $q\bar q$ channel are slightly different for valence and sea quarks.)

We also note that for discussing the size of the power corrections in this context the LO cross section actually provides a better reference value than the full NLO or NNLO corrections. The reason is that the size of the latter relative to the LO cross section can itself strongly depend on the process. Furthermore, at the typical $\tau_\cut$ values of interest the logarithms are so large that they essentially compensate any $\alpha_s$ suppression in the power corrections.
Finally, this allows us to directly compare the numerical size of the corrections for different orders and channels, and also makes it simple to add the channels.

We calculate the full $\Tau_0$ spectrum at $\ord{\alpha_s}$ and $\ord{\alpha_s^2}$ using the $Z+1$-jet (N)LO 
calculation from MCFM8~\cite{Campbell:1999ah, Campbell:2010ff, Campbell:2015qma, Boughezal:2016wmq}, which allows one to generate points down to very small values of $\Tau_0 \simeq 10^{-3}\GeV$.
We then subtract the known singular (leading-power) terms in the $\Tau_0$ spectrum~\cite{Stewart:2009yx, Stewart:2010qs, Berger:2010xi} to obtain the complete nonsingular (subleading-power) contributions,
%%%
\begin{equation} \label{eq:dsigmanons}
\frac{1}{\sigma_\LO}\frac{\df\sigma^\nons}{\df\ln\Tau_0}
= \frac{1}{\sigma_\LO}\frac{\df\sigma}{\df\ln\Tau_0} - \frac{1}{\sigma_\LO}\frac{\df\sigma^{(0)}}{\df\ln\Tau_0}
\,.\end{equation}
%%%
This is done separately for the $\alpha_s$ (NLO) and pure $\alpha_s^2$ (NNLO) contributions and separately for the $q\bar q$ and $qg$ channels.%
\footnote{%
At NLO this distinction is unique, since these are the only two combinations of incoming partons. At NNLO, the $q\bar q$ channel is defined to include all contributions that do not have incoming gluons, while the $qg$ channel also includes all $gg$ contributions. The LL coefficient at subleading power does not depend on this choice of grouping since it is absent for the $gg$ channel.}
The $q\bar q$ channel includes the sum over all flavors, as in $\sigma_\LO$. The $qg$ channel includes the sum of the $qg$ and $gq$ contributions with $q$ summed over all quarks and antiquarks.

By construction the nonsingular cross section starts at subleading power and contains terms of all orders in the power expansion. 
By considering $\df\sigma/\df\ln\Tau_0 = \Tau_0\,\df\sigma/\df\Tau_0$ the leading nonsingular contribution scales $\sim\Tau_0$ and must therefore go to zero for small $\Tau_0$. These are the dominant terms we want to study. Note that computing $\sigma^\nons$ also provides an easy cross check that all leading-power singular terms are correctly calculated and exactly cancel in \eq{dsigmanons}, since any miscancellation would spoil this scaling behaviour and would be immediately visible in the nonsingular data.

Due to the huge numerical cancellations between the full and singular results at small $\Tau_0$, the full result has to be generated with extremely high statistical precision in order to obtain the nonsingular result with sufficiently high precision to allow for precise checks and fits of the subleading contributions.

We perform a standard $\chi^2$ fit to the nonsingular NLO and NNLO data in both channels using the functional forms
%%%
\begin{align} \label{eq:fitfun}
F_\mathrm{NLO}(\tau)
&= \frac{\df}{\df\ln\tau}\Bigl\{ \tau \bigl[(a_1 + b_1 \tau + c_1 \tau^2) \ln\tau
\nn\\ & \quad
+ a_0 + b_0\tau + c_0\tau^2 \bigr] \Bigr\}
\,,\nn\\
F_\mathrm{NNLO}(\tau)
&= \frac{\df}{\df\ln\tau}\Bigl\{\tau \bigl[ (a_3 + b_3 \tau) \ln^3\tau + (a_2 + b_2\tau) \ln^2\tau
\nn\\ & \quad
+ a_1 \ln\tau + a_0 \bigr] \Bigr\}
\,,\end{align}
%%%
with $\tau \equiv \Tau_0/m_Z$.
The coefficients at the same order in $\tau$ tend to be highly correlated, since the different powers of $\ln\tau$ have very similar shapes. To obtain reliable fit results it is thus crucial to ensure that the fit is unbiased. An important consideration is the choice of fit range in $\Tau_0$ and the number of fit coefficients.

Regarding the fitted coefficients, we are ultimately interested in the leading coefficients $a_1$ and $a_0$ at NLO and $a_3$ and $a_2$ at NNLO. When fitting the leading coefficients, neglecting higher-power corrections in the fit (which are of course present in the data) corresponds to a theoretical uncertainty in the fit model. We take this uncertainty into account, with the correct correlations among the bins, by including the higher-power $b_i$ and $c_i$ coefficients as additional nuisance parameters in the fit. With the very precise data needed to get a precise determination of the $a_i$ coefficients, it is essential to do so, because even in a region in $\tau$ where the higher-power contributions might naively seem negligible, they can have a nontrivial influence on the fit as soon as their nominal contribution becomes comparable with the statistical uncertainties in the data. (In other words, the correlated theory uncertainties must be taken into account as soon as they become of similar size to the statistical uncertainties.) At NLO, the data is precise enough to require (or allow) including both $b_i$ and $c_i$ coefficients. At NNLO, we include $b_3$ and $b_2$ since we are interested in unbiased results for $a_3$ and $a_2$. (The NNLO data is not precise enough to require or allow including corresponding $b_1$ and $b_0$ terms.)

Regarding the fit range, in principle the best sensitivity comes from the smallest possible $\tau$ values so we always fit down to the lowest available $\tau$ values. However, the data is much less precise toward smaller $\tau$ values due to the larger numerical cancellations, and much more precise toward larger $\tau$ values. The precision in the fit results thus benefits significantly as the fit range is extended toward large $\tau$, but at the same time is in danger of becoming biased. To achieve a precise but still unbiased fit, we increase the fit range until including an additional data point would reduce the standard $p$-value of the fit. Beyond this point, the $p$-value rapidly deteriorates giving a clear indication that the fit becomes biased and the fit model is not able to describe the data any longer. As a cross check, we also check that including an additional coefficient for the selected fit range does not increase the $p$-value of the fit (while it does so when including the next data point). As further cross checks on the fit results, we divide the
data into two independent subsets and perform separate fits for each subset. We also perform several additional fits with both fewer and more coefficients, using the same procedure to select the fit range in each case, and check that we find compatible fit results.

\begin{table}[h]
\centering
\begin{tabular}{cc|cc}
\hline\hline
\multicolumn{2}{c|}{order and channel} & fitted & calculated
\\ \hline
NLO $q\bar q$ & $a_1$ & $+0.25366\pm 0.00131$ & $+0.25509$
\\
NLO $qg$  & $a_1$ & $-0.27697 \pm 0.00113$ & $-0.27720$
\\ \hline
NNLO $q\bar q$ & $a_3$ & $-0.01112\pm 0.00150$ & $-0.01277$
\\
NNLO $qg$ & $a_3$ & $+0.02373 \pm 0.00247$ & $+0.02256$
\\ \hline\hline
\end{tabular}
\caption{Comparison of fitted and calculated values for the LL coefficients.}
\label{tab:LLresults}
\end{table}

\begin{table}[h]
\centering
\begin{tabular}{cc|c}
\hline\hline
\multicolumn{2}{c|}{order and channel} & fitted
\\ \hline
NLO $q\bar q$ & $a_0$ & $+0.13738 \pm 0.00057$
\\
NLO $qg$ & $a_0$ & $-0.40062\pm 0.00052$
\\ \hline
NNLO $q\bar q$ & $a_2$ & $-0.04662\pm 0.00180$
\\
NNLO $qg$ & $a_2$ & $+0.04234 \pm 0.00242$
\\ \hline\hline
\end{tabular}
\caption{Fit results for the NLL coefficients using the calculated LL coefficients in table~\ref{tab:LLresults} as input.}
\label{tab:NLLresults}
\end{table}

As a check of our calculation we first fit the LL coefficients ($a_1$ at NLO and $a_3$ at NNLO). The results from our default fit for both $q\bar q$ and $qg$ channels are given in table~\ref{tab:LLresults} along with the predicted values from our calculation. In all cases we find excellent agreement. To extract the NLL coefficients ($a_0$ at NLO and $a_2$ at NNLO), we then repeat the fit with the LL coefficients fixed to their predicted values, which allows for a precise determination of $a_0$ and $a_2$, respectively. The results are shown in table~\ref{tab:NLLresults}.
If we approximate the $\xi_a$ and $\xi_b$ dependence of the NLL coefficients in the partonic cross section by the corresponding dependence at LL, we can translate the fitted values for $a_0$ and $a_2$ into the approximate results
%%%
\begin{align}
C_{q\bar q, 0}^{(2,1)}(\xi_a, \xi_b)
&\approx 16.4 \biggl(\delta_a \delta_b+\frac{\delta_a' \delta_b}{2} + \frac{\delta_a \delta_b'}{2}\biggr)
\,, \nn \\
C_{qg, 0}^{(2,1)}(\xi_a, \xi_b)
&\approx -2.45 \,\delta_a \delta_b
\,, \nn \\
C_{q\bar q, 2}^{(2,2)}(\xi_a, \xi_b)
&\approx (378\pm8)  \biggl(\delta_a \delta_b+\frac{\delta_a' \delta_b}{2} +\frac{\delta_a \delta_b'}{2}\biggr)
\,, \nn \\
C_{qg, 2}^{(2,2)}(\xi_a, \xi_b)
&\approx (42.3\pm 0.9)\delta_a \delta_b
\,,\end{align}
%%%
where the uncertainties for the NNLO coefficients arise from the fit uncertainties in the $a_2$.

In \figs{fitNLO}{fitNNLO} we show the nominal fit results for both channels at NLO and NNLO, respectively. The black points show the nonsingular data. The statistical uncertainties are (much) smaller than the size of the data points, except for the lowest points in the NNLO data, where the error bars become visible. (This means that while in all cases the fit quality is good, this fact cannot be judged by eye, so these plots should just be taken as illustration.)
The plots on the left are on a linear scale to show the shape and relative signs of the contributions, while the plots on the right show the same results but taking the absolute value and using a logarithmic scale to highlight the behaviour at small values of $\Tau_0$.
The solid orange line shows the final fit and represents the full nonsingular piece $\sigma^\mathrm{nons}$. The short-dashed extensions show the extrapolation of the fit result beyond the fit range.
To illustrate the contribution from each logarithmic order, the dashed green lines show the LL contributions. The dashed blue lines show the sum of the LL and NLL contributions. In the NLO fits these make up the complete result at $\cO(\tau)$, while at NNLO they lack terms scaling as $\tau\ln\tau$ and $\tau$. In the NNLO fits, the dotted red line shows the full $\cO(\tau)$ contribution including all fitted $a_i$ coefficients (at their central values for illustration only). In both cases, the difference to the full result is due to the higher-power contributions. As expected, the LL and NLL results approach the full nonsingular result toward small values of $\Tau_0$.

In \figs{cumulantNLO}{cumulantNNLO} we show the corresponding results for the missing power corrections $\Delta\sigma(\tau_\cut)$ for $\tau_\cut = \Tau_\cut/m_Z$ at NLO and NNLO respectively, again on a linear scale on the left and logarithmic scale on the right.
In terms of the dimensionless quantity $\tau_\cut$, the results are essentially independent of the precise value of $Q$ (apart from the indirect dependence due to the PDF $x$-range mentioned above and the $\ln Q$ scaling-violation term).
The solid orange, blue dashed, and green dotted lines have the same meaning as in \figs{fitNLO}{fitNNLO}, showing the full nonsingular result together with its NLL and LL approximation, respectively. The solid green lines show the difference between the full nonsingular and its LL approximation, which corresponds to the new missing contribution $\Delta \sigma$ when the LL subleading contribution is included in the subtractions. Similarly, the blue solid line shows the difference between the full nonsingular and its NLL approximation, where the light-blue band shows the effect of the fit uncertainty on the NLL coefficient. This represents the missing power correction $\Delta\sigma$ when the two leading coefficients are included in the subtractions.

At NLO, shown in \fig{cumulantNLO}, we see an improvement by almost an order of magnitude from including each logarithmic order, as expected from our scaling arguments illustrated in \fig{scaling}.
At NNLO, shown in \fig{cumulantNNLO}, for the $qg$ channel we see the expected hierarchy between LL and NLL terms. For the $q\bar q$ channel, the LL and NLL terms happen to largely cancel numerically, such that the full nonsingular correction is accidentally small, and including only the LL subleading contribution actually increases the size of the missing power corrections, while including both terms leads to an improvement (except where the full corrections happens to cross through zero).
Since the contributions in both channels have the opposite sign, when adding the two channels they partially cancel, and they do more so for the higher-logarithmic pieces. As a result, in the total we see a clear hierarchy: Including the LL coefficient yields a substantial improvement by an order of magnitude. Including the NLL contributions leads to some further improvement, although due to the cancellations between the channels the relative uncertainties from the fit are noticeably larger. This motivates an analytic calculation of the NLL term at subleading power.

Overall, we can conclude that for the range of $\tau_\cut$ values of relevance for $N$-jettiness subtractions for color-singlet production, the inclusion of the analytically computed LL power correction significantly improves the numerical behaviour both at NLO and NNLO. We observe a reduction in the error induced due to missing power corrections by about an order of magnitude by including the LL power correction in the subtraction. The inclusion of the NLL contributions, which we are able to extract numerically with good precision, leads to further improvement. This clearly demonstrates how the analytic calculation of subleading power corrections can be used to significantly improve the achievable numerical accuracy and/or the required computational time for the application of $N$-jettiness subtractions.

%%%%%%%%%%%%%%%%%%%%%%%%%%%%%%%%%%%%%%%%%%%%%%%%%%%%%%%%%%%%%%%%%%%%%%%%%%%%%%%%
%\clearpage
\section{Dependence of power corrections on $N$-jettiness definition}
\label{sec:discuss}
%%%%%%%%%%%%%%%%%%%%%%%%%%%%%%%%%%%%%%%%%%%%%%%%%%%%%%%%%%%%%%%%%%%%%%%%%%%%%%%%

\begin{figure*}[t]
\includegraphics[width=\columnwidth]{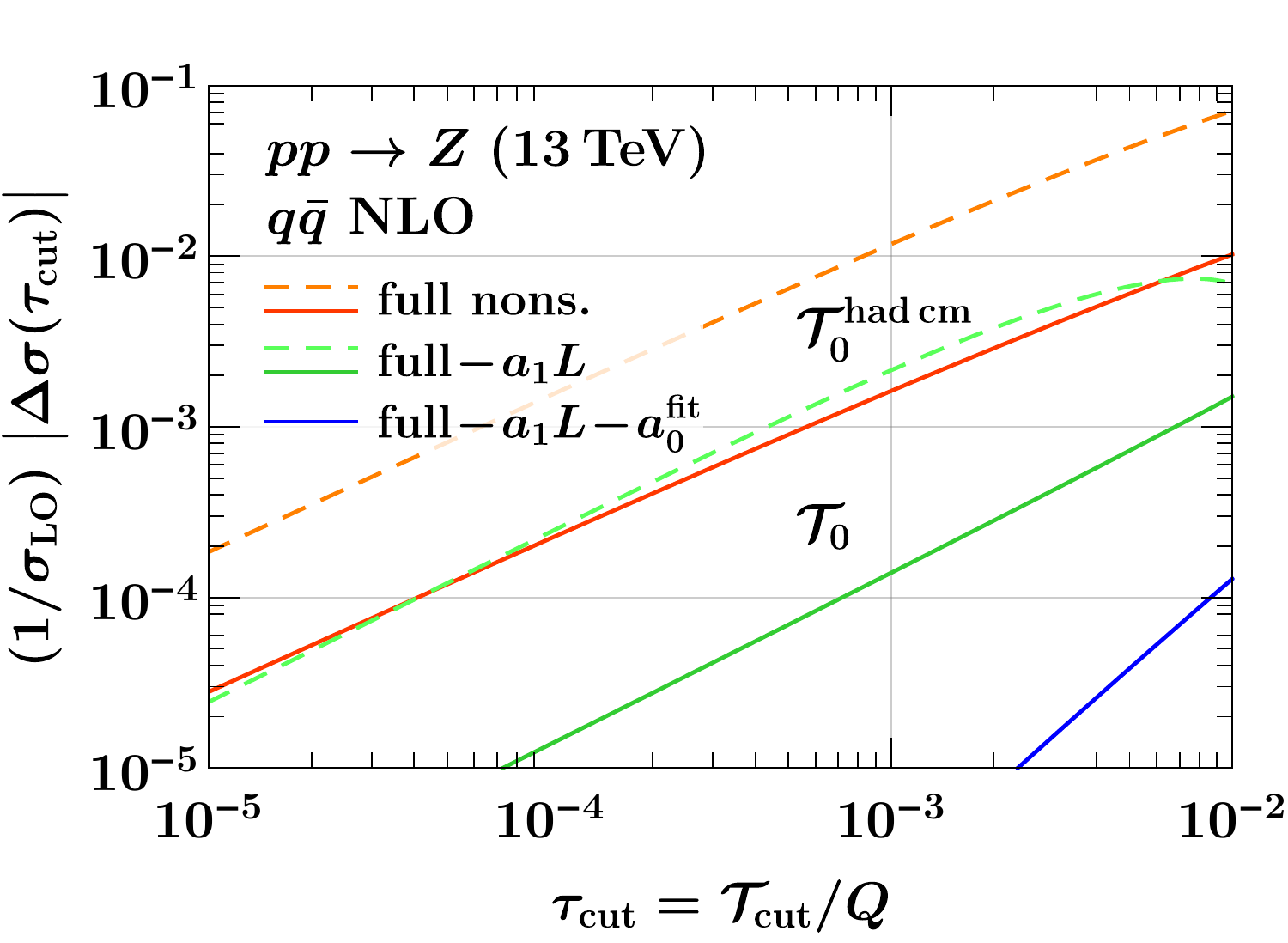}%
\hfill
\includegraphics[width=\columnwidth]{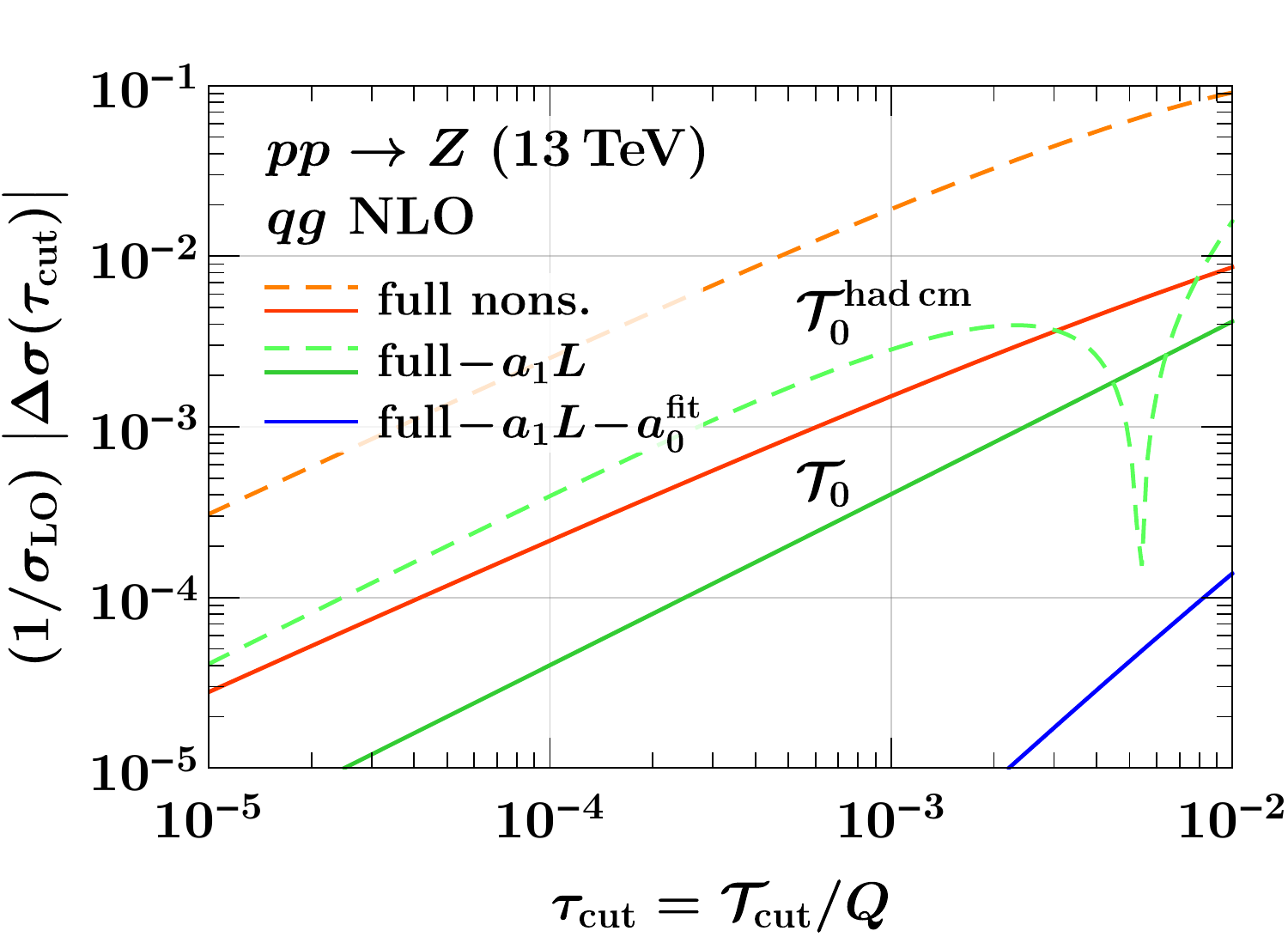}%
\\
\includegraphics[width=\columnwidth]{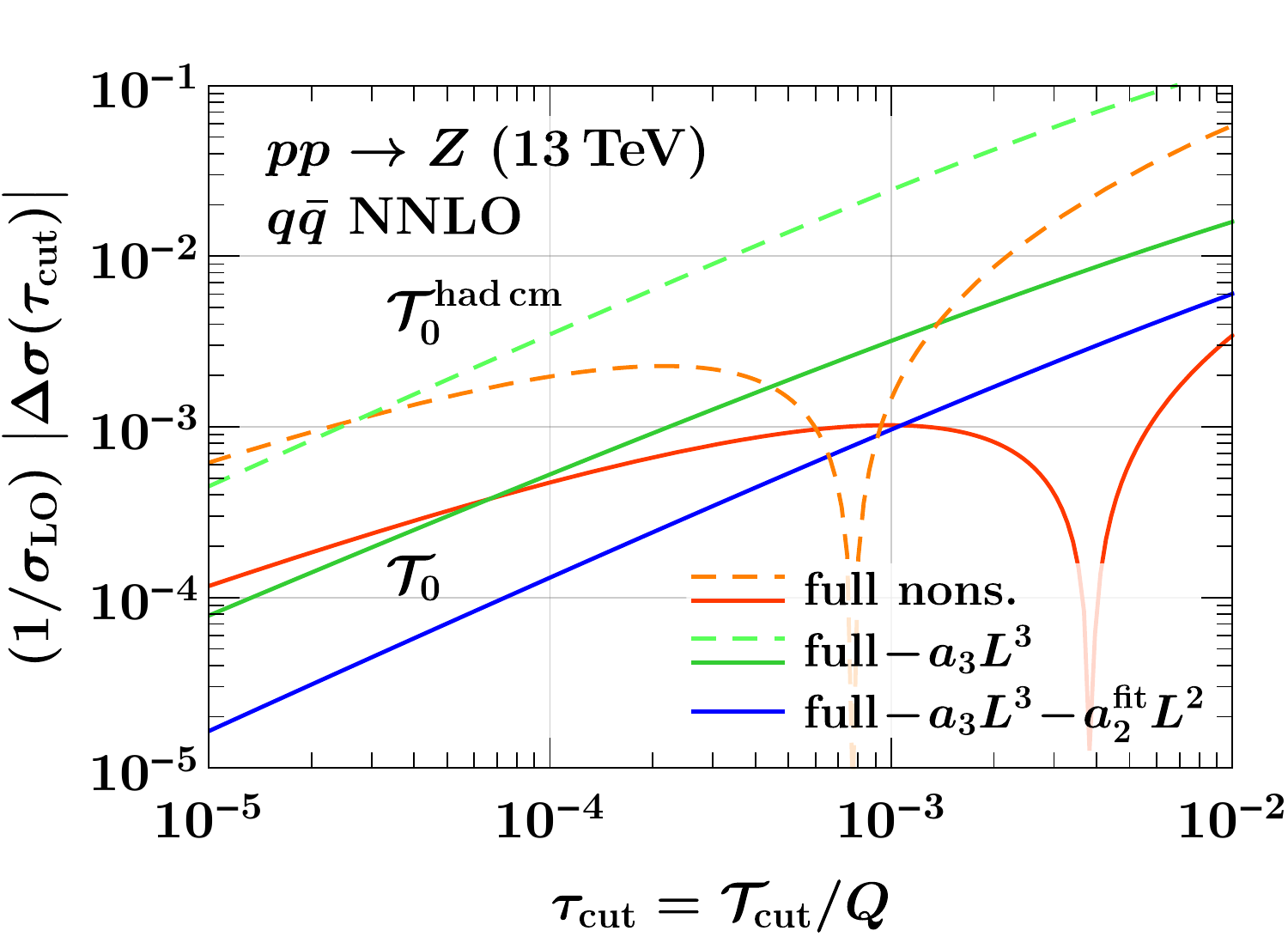}%
\hfill
\includegraphics[width=\columnwidth]{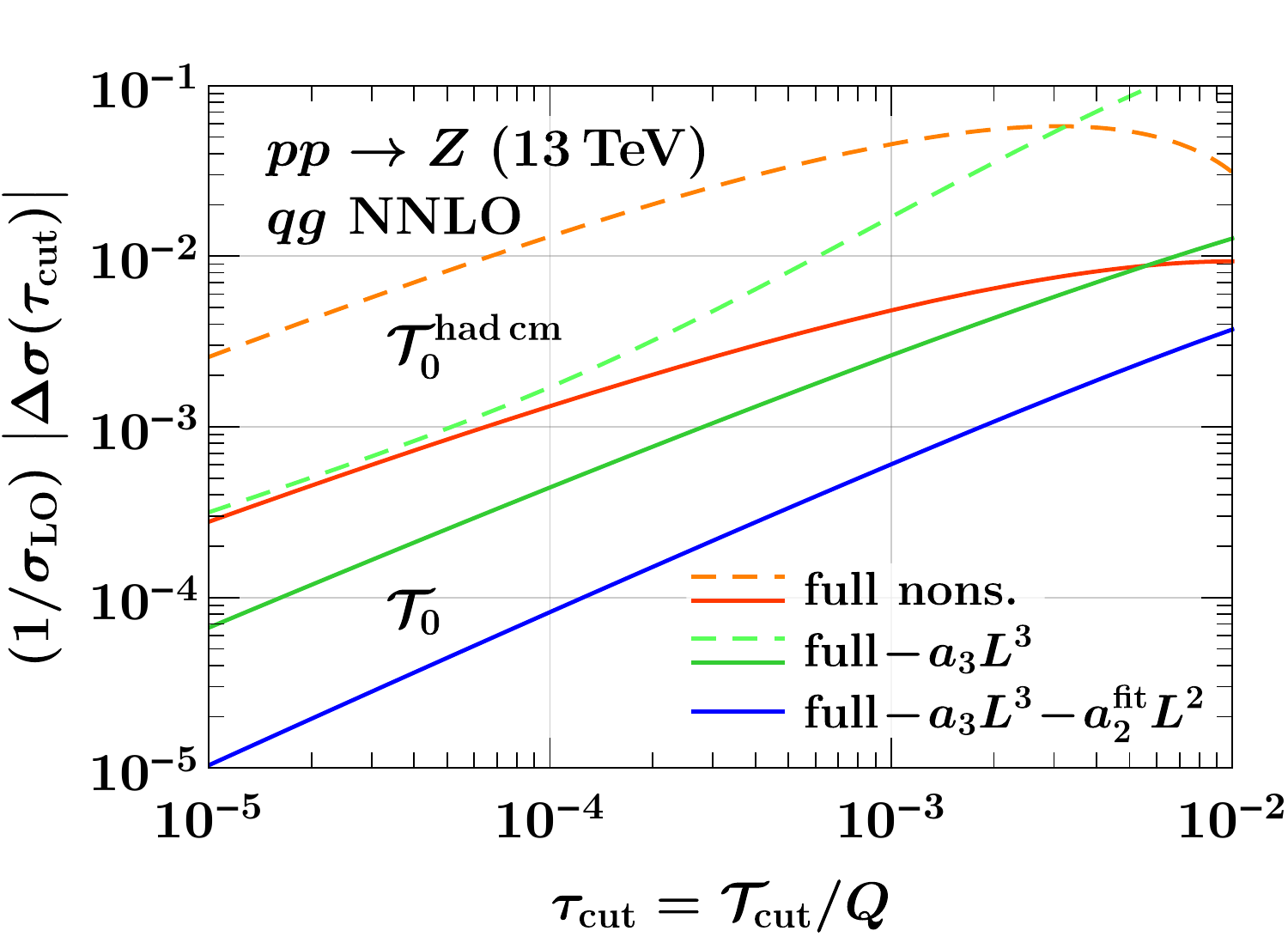}%
\\
\includegraphics[width=\columnwidth]{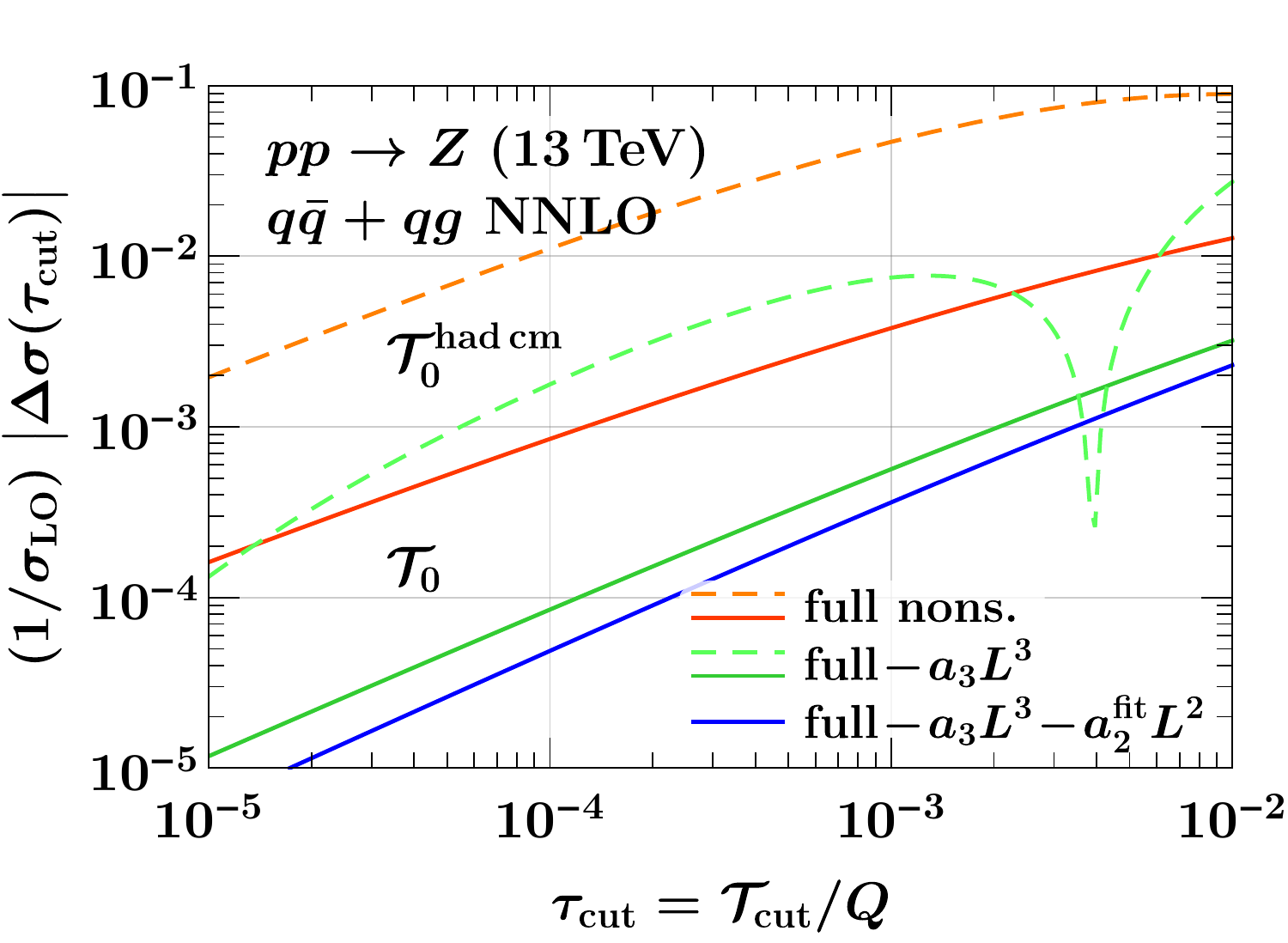}%
%%%
\caption{Comparison of the missing power corrections $\Delta\sigma(\tau_\cut)$ for the hadronic (dashed) and leptonic (solid) definitions of $\Tau_0$. The curves are equivalent to those on the right in \figs{cumulantNLO}{cumulantNNLO}, and a detailed explanation is provided in the text.}
\label{fig:gqNNLO}
\end{figure*}

\begin{figure*}[t]
\includegraphics[width=\columnwidth]{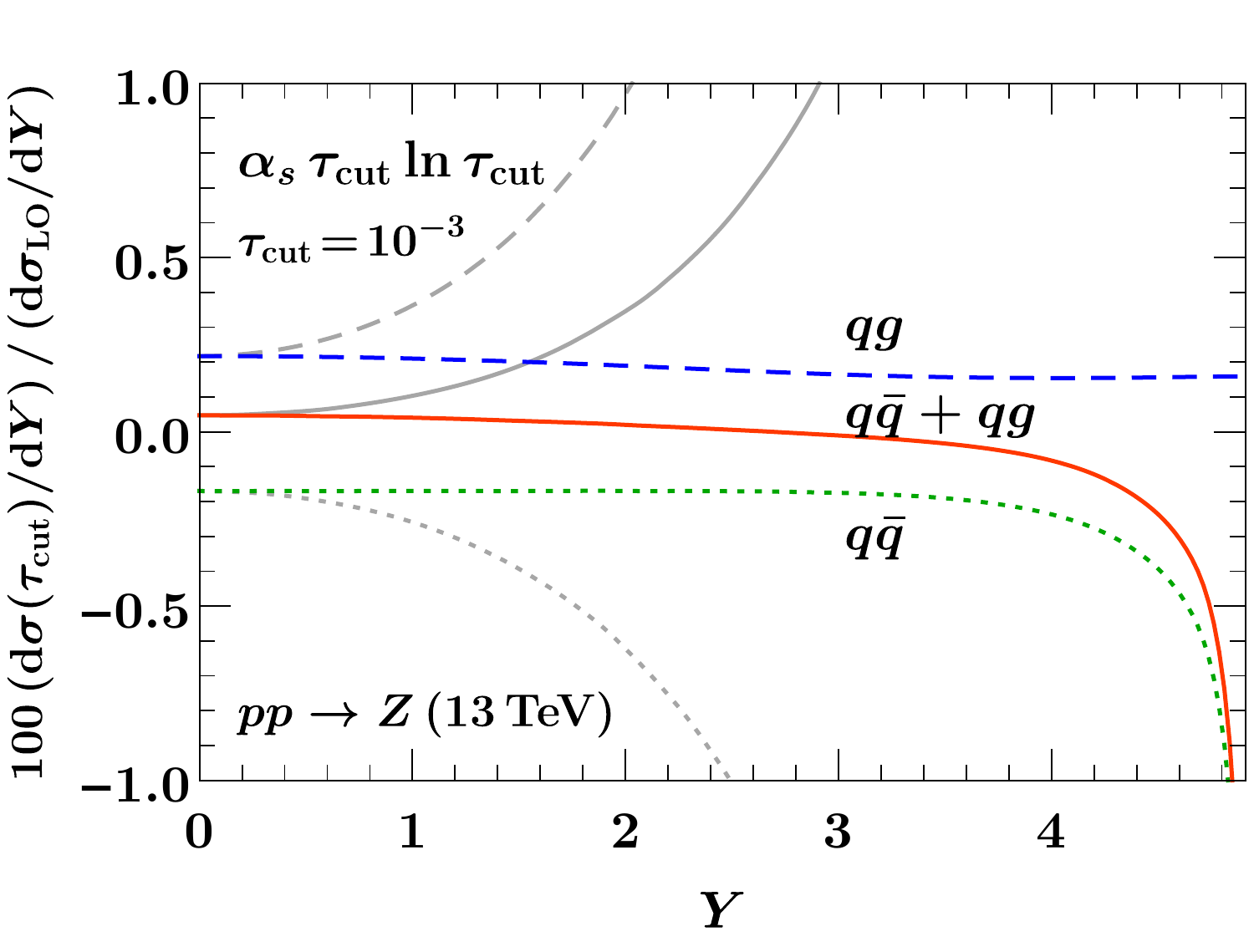}%
\hfill
\includegraphics[width=\columnwidth]{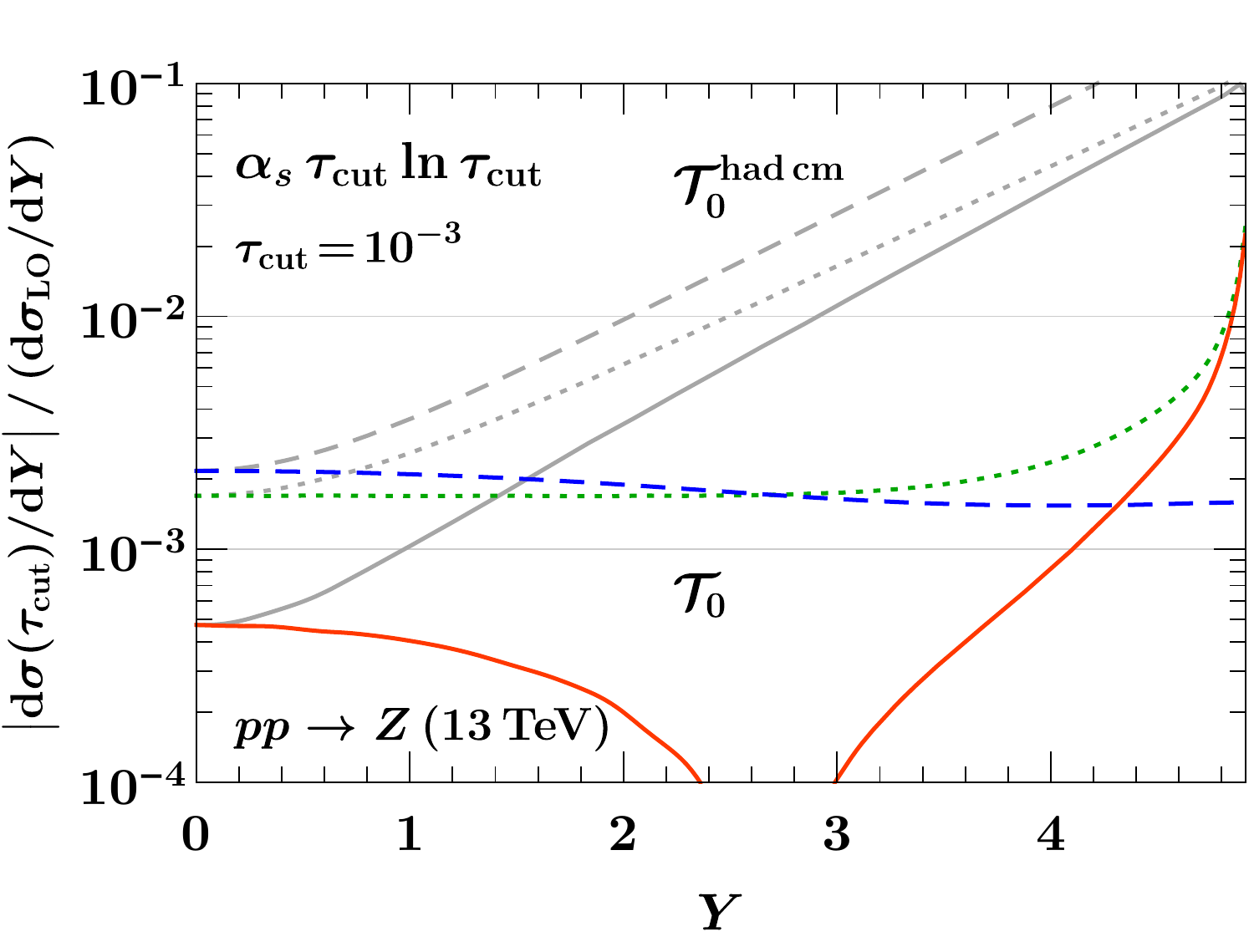}%
\\
\includegraphics[width=\columnwidth]{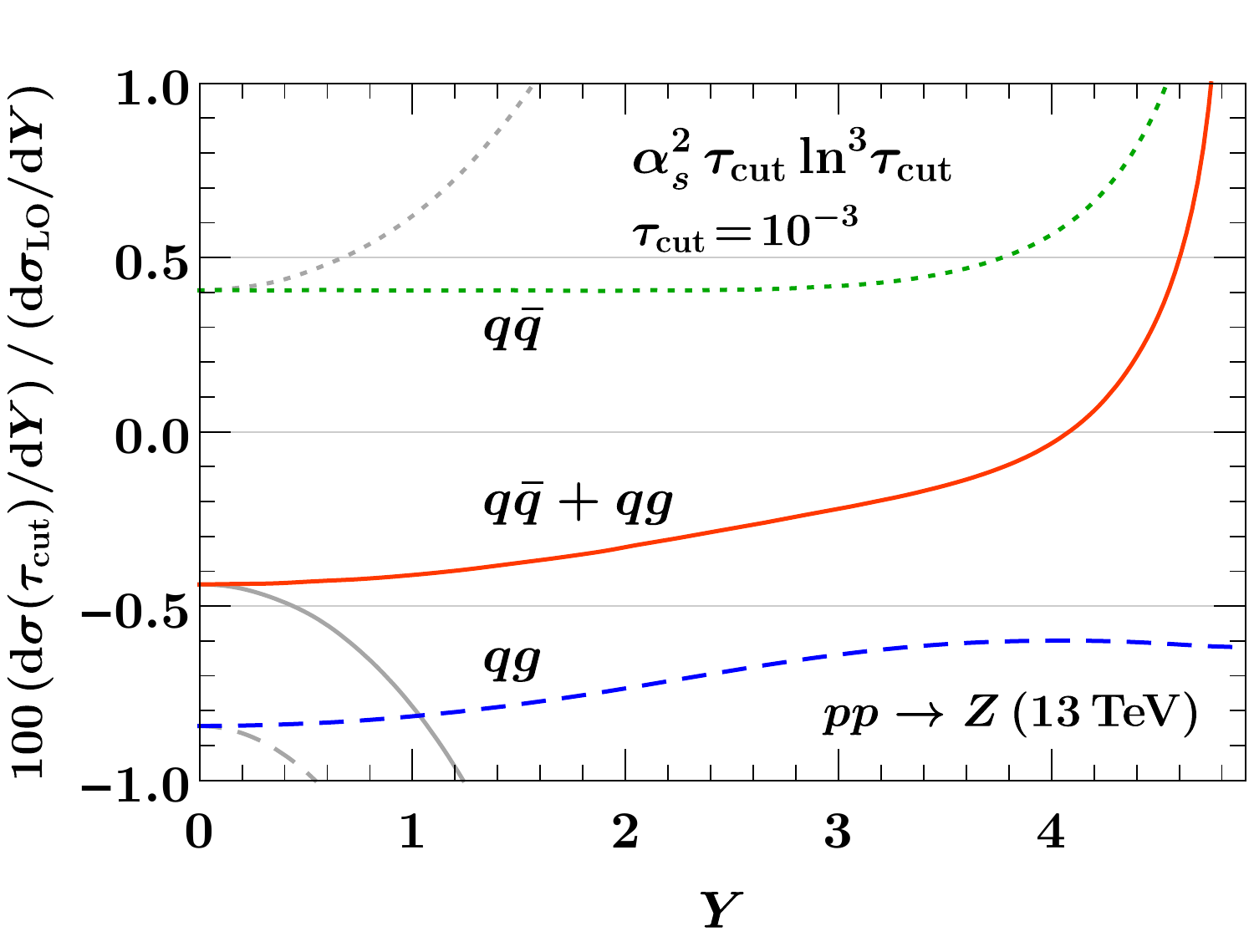}%
\hfill
\includegraphics[width=\columnwidth]{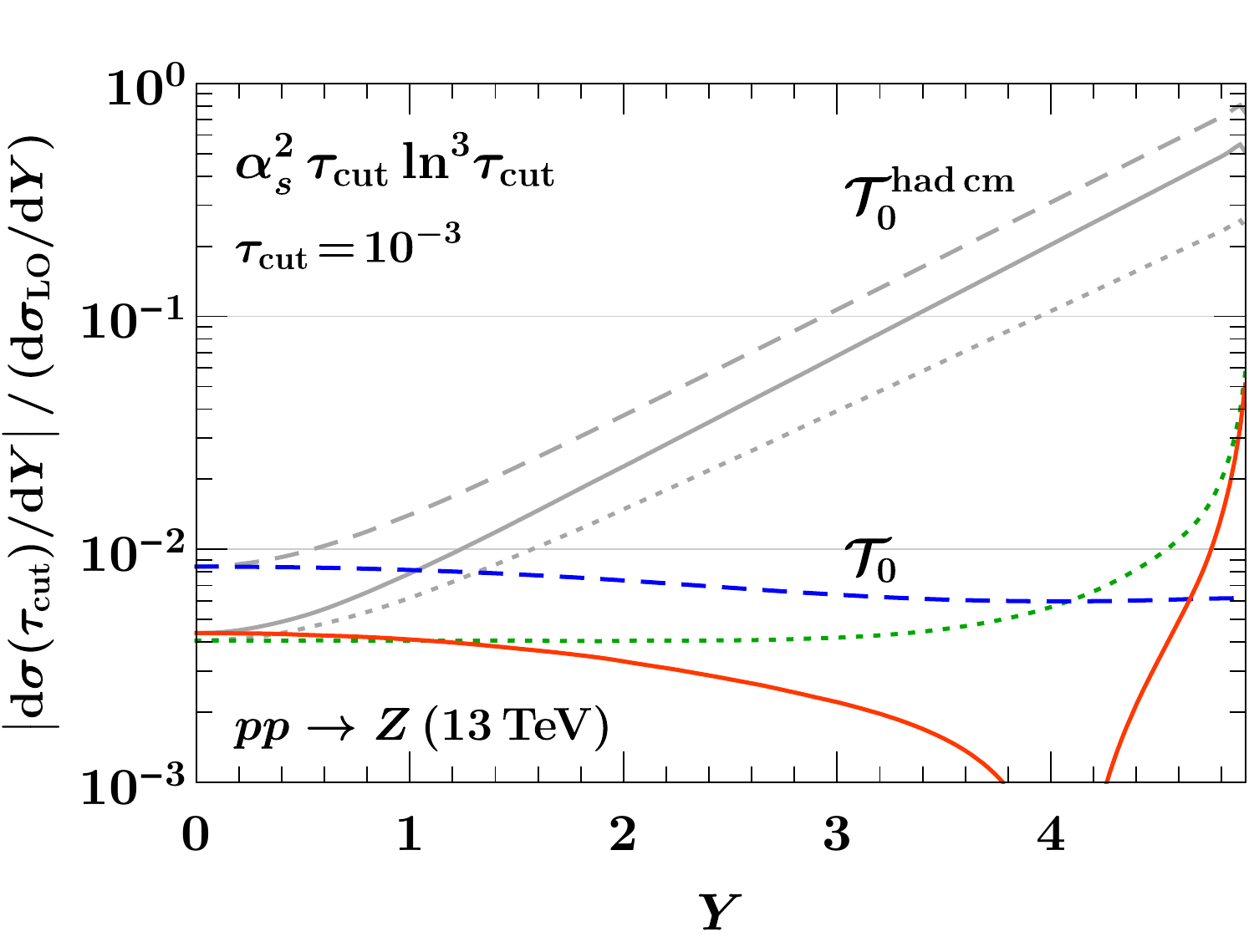}%
%%%
\caption{Leading-logarithmic power correction in the rapidity spectrum at $\ord{\alpha_s}$ (top row) and $\ord{\alpha_s^2}$ (bottom row), using $\tau_\cut = 10^{-3}$. The colored curves show the default definition of $\Tau_0$ in the leptonic frame. The gray curves show the definition  $\Tau_0^\hadcm$ in the hadronic frame, for which the power corrections grow exponentially with rapidity.
The plots on the right are equivalent to those on the left and show the absolute value on a logarithmic scale.
}
\label{fig:rapidity}
\end{figure*}

In this section we discuss how the structure of power corrections depends on the definition of the $N$-jettiness observable, both for the specific case of $0$-jettiness for which we have computed the power corrections analytically, as well as for $N$-jettiness subtractions more generally.

%===============================================================================
\subsection{\boldmath $0$-jettiness}
%===============================================================================

In \subsec{beam_thrust} we considered the leptonic definition of $N$-jettiness, which explicitly accounts for the boost of the Born system due to the $e^Y$ factor in the measure. As shown by the results in \eqs{NLOResult}{NNLOResult}, the power corrections in this case are independent of the Born kinematics, up to the dependence on the PDFs, the numerical impact of which we will study in this section.

It is also interesting to consider the hadronic definition of beam thrust, $\Tau_0^\hadcm$ given in \eq{Njet_def}, which has been used in some applications of $N$-jettiness subtractions. The calculation proceeds identically to the case of the leptonic definition, $\Tau_0$, with only a difference in the measurement function. For $\Tau_0^\hadcm$, we find at NLO
%%%
\begin{align} \label{eq:NLOResulthad}
\widetilde C_{q\bar q, 1}^{(2,1)}(\xi_a, \xi_b)
&= 4C_F \Bigl[ e^Y \delta_a (\delta_b + \delta'_b) + e^{-Y} (\delta_a + \delta_a')\delta_b \Bigr]
 \,, \nn \\
\widetilde C_{qg, 1}^{(2,1)}(\xi_a, \xi_b)
&= -2 T_F \,e^Y\delta_a \delta_b
 \,, \nn \\
\widetilde C_{gq, 1}^{(2,1)}(\xi_a, \xi_b)
&= -2 T_F \,e^{-Y} \delta_a \delta_b
\,,\end{align}
%%%
and at NNLO
%%%
\begin{align} \label{eq:NNLOResulthad}
\widetilde C_{q\bar q, 3}^{(2,2)}(\xi_a, \xi_b)
&= -16 C_F^2 \Bigl[ e^Y \delta_a (\delta_b + \delta'_b) + e^{-Y} (\delta_a + \delta_a')\delta_b \Bigr]
, \nn \\
\widetilde C_{qg, 3}^{(2,2)}(\xi_a, \xi_b)
&=  4 T_F(C_F+C_A)e^Y \delta_a \delta_b
\,, \nn \\
\widetilde C_{gq, 3}^{(2,2)}(\xi_a, \xi_b)
&=  4 T_F(C_F+C_A)e^{-Y} \delta_a \delta_b
\,. \end{align}
%%%
Here, the $\mathcal{O}(\tau^\hadcm)$ power corrections grow like $e^{\pm Y}$, and are thus exponentially enhanced at forward rapidities. The reason for this can be easily understood as follows. In SCET, the collinear and soft particles are assigned a scaling $p_c=(p_c^-,p_c^+,p_c^\perp)\simeq (1,\lambda^2,\lambda)$ and $p_s \simeq (\lambda^2, \lambda^2, \lambda^2)$, and the power expansion is in terms of $\lambda^2\simeq p_c^+/p_c^- \simeq p_s/p_c^-$. For the leptonic definition, it is effectively performed in the frame where the Born system is at rest, and where the large minus momentum of the colliding partons are fixed to $\hat p_a^- = \hat p_b^+ = Q$. At the same time, $\Tau_0$ scales as $\Tau_0 \simeq \min\{\hat p_c^+, \hat p_c^-\}$, so the effective expansion parameter is $\lambda^2 \simeq \Tau_0/Q$.
In contrast for the hadronic definition, the expansion is effectively performed in the hadronic frame and is in terms of $\Tau_0^\hadcm/p_a^- = e^Y\Tau_0^\hadcm/Q$ or $\Tau_0^\hadcm/p_b^+ = e^{-Y}\Tau_0^\hadcm/Q$. This is precisely what leads to the $e^{\pm Y}$ factors in \eqs{NLOResulthad}{NNLOResulthad}. This also means that the power expansion for $\Tau_0^\hadcm$ actually significantly deteriorates at forward rapidities, with the $n$th-order power corrections scaling as $e^{nY}(\tau^\hadcm)^n$. (This deterioration causes power corrections in $\Tau_0^\hadcm$ to be much larger than those for $\Tau_0$, but does not change the fact that for both variables  in the small $\tau$ region the nonsingular corrections are still much smaller than the leading power singular corrections.)

The sensitivity of the power correction to the boost of the partonic frame can also be understood physically by considering a soft emission from the incoming collinear partons. At leading power, this is described by the eikonal matrix element, which is independent of the large momentum fraction of the collinear parton. At subleading power, this is no longer true, and the matrix element depends explicitly on the momentum fraction of the collinear parton.

We have also performed our numerical analyses of the previous section for $\Tau_0^\hadcm$. The much larger size of the nonsingular corrections are clearly visible in the nonsingular data. However, the fit range cannot be extended above $\Tau_\cut = 10^{-2}\GeV$, beyond which the power expansion deteriorates. As a result, we are only able to obtain a qualitative cross check for the leading coefficients. Fixing the leading coefficient to the analytic prediction, the fit provides a good interpolation of the nonsingular data, but we were not able to extract a meaningful result for the NLL coefficients.

In \fig{gqNNLO}, we compare the leptonic (solid) and hadronic (dashed) definitions of $0$-jettiness, showing the cumulant $\sigma^\nons(\tau_\cut)$ as a function of $\tau_\cut = \Tau_0^{(\hadcm)}/Q$ for the $q\bar q$ and $qg$ channels at NLO and NNLO. The orange curves show the full nonsingular result, while the green curves show the result removing the LL contribution. For the leptonic definition we also show the result removing the fitted NLL contributions in blue, i.e., the solid curves are equivalent to those in \figs{cumulantNLO}{cumulantNNLO}.
In all cases the power corrections are about an order of magnitude larger for $\Tau_0^\hadcm$ than for $\Tau_0$, independently of $\tau_\cut$ (except in the region where the result happens to cross through zero). In fact, in many cases, the power corrections for $\Tau_0^\hadcm$ even after removing the leading-power contribution is larger than the full nonsingular leptonic result.
Thus by switching from the hadronic definition to the leptonic definition and  including the analytic calculation of the leading power correction, one can gain an improvement in the precision of the $N$-jettiness subtractions by two orders of magnitude.

To illustrate the rapidity dependence, in \fig{rapidity} we show the size of the leading-logarithmic power corrections as a function of the vector-boson rapidity (using a fixed value $\tau_\cut = 10^{-3}$) on a linear and logarithmic scale. The results for $\Tau_0$ are shown in dotted green for the $q\bar q$ channel, dashed blue for the $qg$ channel, and solid orange for their sum. For comparison, the corresponding results for $\Tau_0^\hadcm$ with the same dashing are shown in grey. As expected, the power corrections for $\Tau_0$ are essentially flat in rapidity, except for the $q\bar q$ channel at the very endpoint due the sensitivity of the contributions involving PDF derivatives to the PDF endpoint. In contrast, for $\Tau_0^\hadcm$ the exponential growth of the power corrections with rapidity is clearly seen. This explains the numerical behavior that was observed in Ref.~\cite{Boughezal:2016wmq}, where the convergence of the $N$-jettiness subtractions strongly depended on the application of rapidity cuts. On the other hand, in Ref.~\cite{Gaunt:2015pea} where the leptonic definition was used, the rapidity spectrum was completely well-behaved for any $Y$.\footnote{%
Recently, the $pp\to H+1$-jet NNLO calculation in Ref.~\cite{Chen:2016zka} found a disagreement with the results in Ref.~\cite{Boughezal:2015aha} based on $1$-jettiness subtractions, amounting to a $\simeq 30\%$ difference in the NNLO coefficient. The $\Tau_\cut$ values used in Ref.~\cite{Boughezal:2015aha} were between $0.05-0.1\GeV$ which for typical $Q \simeq 150-200\GeV$ corresponds to $\tau_\cut \gtrsim 3\times 10^{-4}$. From our scaling estimates and the fact that Ref.~\cite{Boughezal:2015aha} appears to use a hadronic-frame definition for $\Tau_1$, for which one can expect the power corrections to be enhanced, it is possible that this size of difference in the NNLO contribution could be caused by the power corrections.}

Since the purpose of the $N$-jettiness subtractions is to perform NNLO calculations fully differential in the Born phase space, it is of course quite undesirable to have a strong dependence of the size of missing power corrections on the Born phase space as is the case for $\Tau_0^\hadcm$.
Hence, a beneficial feature of the leptonic definition is that the relative size of missing power corrections and thereby the accuracy of the subtraction method is essentially independent of the Born phase space and thus also of additional Born level cuts. In practice this means that the numerical accuracy does need to be reevaluated on a case-by-base basis depending on the applied cuts.

%===============================================================================
\subsection{\boldmath $N$-jettiness}
%===============================================================================

The above considerations regarding the size of power corrections are not limited to the case of $q \bar q$-initiated color-singlet production and can be applied more generally to ensure a well-behaved power expansion.

Since $N$-jettiness subtractions are applied at the level of a theoretical calculation, it is always possible to use a definition of $N$-jettiness that incorporates the boost of the Born system relative to the frame of the hadronic collision, as discussed in Ref.~\cite{Stewart:2010tn},
%%%
\begin{equation}\label{eq:TauN_def}
\Tau_N = \sum_k \min_i \Bigl\{ \frac{2 q_i\cdot p_k}{Q_i} \Bigr\}
\,,\end{equation}
%%%
where the minimum runs over $i = \{a, b, 1, \ldots, N\}$.
In principle, more general $N$-jettiness measures $d_i(p_k)$ are possible as well. The above
choice $d_i(p_k) = (2q_i\cdot p_k)/Q_i$ is convenient for theoretical calculations,
because it is linear in the momenta $p_k$~\cite{Jouttenus:2011wh, Jouttenus:2013hs}.
The $q_i$ are massless reference momenta corresponding to the momenta of the hard partons
present at Born level,
%%%
\begin{equation}
q_i^\mu = E_i n_i^\mu
\,,\qquad
n_i^\mu = (1, \vec n_i)
\,,\qquad
\abs{\vec n_i} = 1
\,.\end{equation}
%%%
In particular, the reference momenta for the incoming partons are given by
%%%
\begin{equation}
q_{a,b}^\mu = x_{a,b} \frac{E_{\rm cm}}{2}\, n^\mu_{a,b}
\,,\qquad
n_{a,b}^\mu = (1, \pm \hat z)
\,,\end{equation}
%%%
where
%%%
\begin{align} \label{eq:beamref}
2E_a &= x_a E_{\rm cm} = n_b \cdot (q_1 + \cdots + q_N + q_L) = Q e^Y
\,, \nn \\
2E_b &= x_b E_{\rm cm} = n_a \cdot (q_1 + \cdots + q_N  + q_L) = Q e^{-Y}
\,, \nn \\
Q^2 &= x_a x_b E_{\rm cm}^2
\,, \qquad
Y = \frac{1}{2}\ln\frac{x_a}{x_b}
\,.\end{align}
%%%
Here, $q_L$ is the total momentum of any additional color-singlet particles in the Born process,
and $Q$ and $Y$ now correspond to the total invariant mass and rapidity of the Born system.
A more detailed discussion of the construction of the $q_i$ in the context of fixed-order calculations and $N$-jettiness subtractions can be found in Ref.~\cite{Gaunt:2015pea},

The measure factors $Q_i$ influence the singular behaviour of the $\Tau_N$ cross section
and therefore also the power expansion. Several choices have been discussed in Ref.~\cite{Jouttenus:2011wh, Jouttenus:2013hs}. The invariant-mass measure is given by choosing a common $Q_i = Q$,
and this automatically incorporates correctly the boost $Y$ of the Born system via \eq{beamref}.
A class of geometric measures is obtained by choosing $Q_i = \rho_i\,2 E_i$, where
the $\rho_i$ are dimensionless. This is convenient in that it makes $\Tau_N$ independent
of the energies $E_i$ and only dependent
on the directions $n_i$. In this case, a simple way to correctly incorporate the boost $Y$
is to choose $\rho_i \equiv 1$ in the Born frame, i.e., the frame where the Born system has $Y = 0$,
such that
%%%
\begin{equation}\label{eq:TauN_Born}
\Tau_N = \sum_k \min_i \bigl\{ \hat n_i\cdot \hat p_k \bigr\}
\,,\end{equation}
%%%
where $\hat p_k$ and $\hat n_i$ are the final-state momenta and reference
directions in the Born frame. For the beam contributions, this reduces to
the leptonic definition $\min\{\hat p_k^+, \hat p_k^-\}$. In the hadronic frame
this corresponds to choosing $\rho_{a,b} = e^{\pm Y}$, while the $\rho_{i \geq 1}$
are more complicated.

By the same scaling arguments as for $\Tau_0$, taking into account the boost $Y$ in the definition of $\Tau_N$ as described above, one ensures that the power expansion is well behaved, and that power corrections do not grow exponentially with $Y$, which would be the case for example if $\Tau_N$ was defined as in \eq{TauN_Born}, but in the hadronic frame.
In particular, this applies to processes such as $pp\to$ $W/Z/H+$ jet, to which $N$-jettiness subtractions have already been applied.  This definition also applies to processes involving only jets at Born level, such as $pp\to$ dijets which is an obvious next target for the application of $N$-jettiness subtractions.
Analytic calculations of the leading power corrections can then be used for further improvements, as they become available.
It would also be interesting to study the definition of the $N$-jettiness axes for the final-state jets in more detail, to understand if they can be chosen in such a manner as to further reduce the power corrections.

%%%%%%%%%%%%%%%%%%%%%%%%%%%%%%%%%%%%%%%%%%%%%%%%%%%%%%%%%%%%%%%%%%%%%%%%%%%%%%%%
\section{Conclusions}
\label{sec:conclusions}
%%%%%%%%%%%%%%%%%%%%%%%%%%%%%%%%%%%%%%%%%%%%%%%%%%%%%%%%%%%%%%%%%%%%%%%%%%%%%%%%

We have presented an analytic calculation of the $\alpha_s \ln\tau$ and $\alpha_s^2\ln^3\!\tau$ power corrections for thrust in $e^+e^-$ collisions and $0$-jettiness (beam thrust) for $q\bar q$ annihilation in $pp$ collisions. We showed how subleading power
corrections for event shape observables can be systematically computed using SCET, and derived general consistency relations of the subleading coefficients from the cancellation of $1/\eps$ poles.
We have checked our analytic expressions by comparing with numerical fixed-order results, as well as analytically, by comparing with the known NNLO result in the threshold limit, finding excellent agreement in all cases.

Our calculation clearly shows the advantage of using SCET for calculating power corrections.
Besides a systematic organization of operators and Lagrangian
insertions, the consistency relations in \eq{constraint_setup} lead to a significant simplification and a valuable cross check of the calculation. Already at NLO, the SCET calculation
contains divergences in both the collinear and soft matrix element, whose cancellation leads to the nontrivial constraint in \eq{one_loop_constraint}. In contrast, the full NLO QCD calculation does not have $1/\epsilon$ poles and a priori
does not provide similar constraints.
At higher orders, the importance of the consistency constraints becomes evident, as they lead to a significant reduction in the number of matrix elements that are needed for the leading and subleading logarithms, as shown in \eq{constraints_final}. Furthermore, checking the relations in \eq{constraints_summary} provides an important cross check
of our calculation.

Our analytic results have a number of interesting features, including the appearance of a $C_A$ color structure in the coefficient of the $\alpha_s^2 \ln^3\!\tau$ term in the $q g$ channel. This color structure arises from a limit in which a quark becomes soft, or two quarks become collinear, which does not have an analog at leading power.
It would be interesting to understand this structure also at higher orders in $\alpha_s$, and how it arises
from the renormalization group evolution at subleading power.

Our computation of the leading power corrections allows for the improvement of the $N$-jettiness subtractions for Drell-Yan-like
color-singlet production. Including the LL power correction in the subtractions reduces the error due to missing power corrections by about an order of magnitude. The analytic calculation of the LL contributions also allowed us to numerically extract the NLL contributions, further reducing the error. We find that these contributions are desirable to have a stable reduction in all channels, emphasizing the importance of computing the NLL coefficient analytically.
Our numerical results also confirm well the naive scaling estimates for their size.
Moreover, the explicit knowledge of the dominant power corrections allows for an \emph{a priori}
determination of the error induced by missing power corrections.

Finally, we have emphasized the importance of the definition of the $N$-jettiness variable for ensuring a well-behaved power expansion, which applies to any $N$. In particular, we found that defining $N$-jettiness in the hadronic center-of-mass frame artificially induces power corrections that grow exponentially with rapidity. On the other hand, this does not happen when taking into account the boost of the Born system in the definition of $N$-jettiness (as was done in the original definitions of beam thrust and $N$-jettiness~\cite{Stewart:2009yx, Stewart:2010tn}).
This stability of the power corrections with rapidity is important for the applicability of the $N$-jettiness subtractions for computing arbitrary differential distributions.

There are a number of interesting directions that we plan to address in future work. While we have focused here on the power corrections for $q \bar{q}$-initiated Drell-Yan-like processes, the same approach can be used to calculate the power corrections for $gg$-initiated color-singlet production, as relevant for gluon-fusion Higgs production.
It would also be interesting to extend our calculation to subleading logarithms, which would be facilitated by exploiting the consistency relations, and it would be interesting to derive renormalization group equations to predict the series of logarithms at subleading power. Finally, a key feature of the $N$-jettiness subtraction method is that it extends to processes involving final-state jets. It will be important to extend our results to calculate power corrections for $\Tau_N$ with $N>0$, allowing for analytic control over the power corrections for general NNLO $N$-jettiness subtractions.

%{\bf EDIT FOR CONCLUSION:} In addition to their application to $N$-jettiness subtractions, subleading power corrections are also important in the context of
%high-precision resummation. While the most singular terms in the cross section, $\df\sigma^{(0)}/\df\tau_N$, are well understood for a wide range of observables, and can be resummed to high perturbative orders, much less is known about the resummation structure of power corrections. This is interesting both theoretically for understanding the soft and collinear limits of gauge theories, as well as for explicit computations, where the inclusion of power corrections could significantly reduce the uncertainties in combining resummed calculations with full fixed-order results. This would require
%a complete subleading factorization theorem(s) and associated renormalization group evolution, which is beyond the
%scope of this paper. The subleading fixed-order expansion are an important first step in this direction.

%%%%%%%%%%%%%%%%%%%%%%%%%%%%%%%%%%%%%%%%%%%%%%%%%%%%%%%%%%%%%%%%%%%%%%%%%%%%%%%%
\begin{acknowledgments}

We thank the Erwin Schr\"odinger Institute and the organizers of the ``Challenges and Concepts for Field Theory and Applications in the Era of LHC Run-2'' workshop for hospitality and support while portions of this work were completed.
This work was supported in part by the Office of Nuclear Physics of the U.S.
Department of Energy under Contract No. DE-SC0011090, by the Office of High Energy Physics of the U.S. Department of Energy under Contract No. DE-AC02-05CH11231,
by the DFG Emmy-Noether Grant No. TA 867/1-1,
the Simons Foundation Investigator Grant No. 327942, and the LDRD Program of LBNL.
While this work was in progress we also became aware of calculations of power suppressed terms that are being carried out by R.~Boughezal, X.~Liu and F.~Petriello.

\end{acknowledgments}
%%%%%%%%%%%%%%%%%%%%%%%%%%%%%%%%%%%%%%%%%%%%%%%%%%%%%%%%%%%%%%%%%%%%%%%%%%%%%%%%

\bibliography{subleading.bib}

\end{document}